\documentclass[fleqn,usenatbib]{mnras}
\usepackage{newtxtext,newtxmath}
\usepackage[T1]{fontenc}
\usepackage{ae,aecompl}
\usepackage{graphicx}	
\usepackage{graphics}
\usepackage{threeparttable}
\usepackage[normalem]{ulem}
\usepackage{bm}

\newcommand{\msun}{\mbox{M$_{\odot}$}}
\newcommand{\fss}{\hbox{$.\!\!^\mathrm{s}$}}        
\newcommand{\h}{$^\mathrm{h}$}
\newcommand{\m}{$^\mathrm{m}$}

\newcommand{\nh}{$N_\mathrm{H}$}
\newcommand{\gr}{$\gamma$-ray}
\newcommand{\psr}{J1957}
\newcommand{\jpsr}{J1957+5033}
\newcommand{\ergs}{erg~s$^{-1}$}
\newcommand{\flux}{erg~s$^{-1}$~cm$^{-2}$}
\newcommand{\phs}{ph~cm$^{-2}$~s$^{-1}$~keV$^{-1}$}
\newcommand{\xmm}{\textit{XMM-Newton}}
\newcommand{\chan}{\textit{Chandra}}
\newcommand{\fermi}{\textit{Fermi}}
\newcommand{\degs}{\ifmmode ^{\circ}\else$^{\circ}$\fi}
\newcommand{\amin}{\ifmmode ^{\prime}\else$^{\prime}$\fi}
\newcommand{\tc}{t_\mathrm{c}}
\newcommand{\Teff}{T_\mathrm{eff}}
\newcommand{\Ts}{T_\mathrm{s}}

\title[Middle aged $\gamma$-ray pulsar J1957+5033 in X-rays]{Middle aged $\gamma$-ray pulsar J1957+5033 in X-rays: pulsations, thermal emission and nebula} 

\author[Zyuzin et al.]{D. A. Zyuzin$^{1}$\thanks{E-mail: da.zyuzin@gmail.com},
A. V. Karpova$^{1}$,
Y. A. Shibanov$^{1}$,
A. Y. Potekhin$^{1}$,
V. F. Suleimanov$^{2,3,4}$
\\
$^{1}$Ioffe Institute, Politekhnicheskaya 26, St. Petersburg, 194021, Russia\\
$^{2}$Institut f\"{u}r Astronomie und Astrophysik, Sand 1, D-72076 T\"{u}bingen, Germany\\
$^{3}$Kazan (Volga region) Federal University, Kremlevskaja str., 18, Kazan 420008, Russia \\
$^{4}$Space Research Institute of the Russian Academy of Sciences, Profsoyuznaya Str. 84/32, Moscow 117997, Russia
}

\date{Accepted XXX. Received YYY; in original form ZZZ}

\pubyear{2020}
\begin{document}
\label{firstpage}
\pagerange{\pageref{firstpage}--\pageref{lastpage}}
\maketitle

\begin{abstract}
We analyze new 
\xmm\ and archival \chan\ observations of the middle-aged $\gamma$-ray radio-quiet pulsar \jpsr.
 We detect, for the first time, X-ray pulsations with the pulsar spin period of the point-like source coinciding by position with the  pulsar. This 
 confirms the pulsar nature of the source.   
In the 0.15--0.5 keV band, there is a single pulse per period and the pulsed fraction  is  $\approx18\pm6$ per cent. 
In this band, 
the pulsar spectrum is dominated by a thermal emission component that likely 
comes from the entire  surface of the neutron star, while at higher energies ($\gtrsim0.7$ keV) it is described by a power law with the photon index 
$\Gamma \approx 1.6$. We construct new hydrogen atmosphere models for neutron stars with dipole magnetic fields and non-uniform surface 
temperature distributions  with relatively low effective temperatures. 
We use them in the spectral analysis and 
derive the pulsar average effective 
temperature of $\approx(2-3)\times10^5$~K. This
makes \jpsr\ the coldest among all known thermally emitting neutron stars with ages below 1 Myr. Using the interstellar extinction--distance relation, we constrain the distance to the pulsar in the range of 0.1--1 kpc. 
We compare the obtained X-ray thermal luminosity with those for other neutron stars and 
various neutron star cooling models and set some constraints on latter. 
We observe a faint trail-like feature, elongated $\sim 8$ arcmin 
from \jpsr.  
Its spectrum can be described by a power law with a photon index $\Gamma=1.9\pm0.5$  
suggesting that it is likely a pulsar wind nebula powered  by \jpsr.

\end{abstract}



\section{Introduction}
\label{sec:intro}

Neutron stars (NSs) are born in supernova explosions at very high temperatures $\sim 10^{11}$ K \citep[e.g.][and references therein]{Mueller20}.
They lose the initial thermal energy via neutrino emission from their interiors and then via photon emission from their surfaces
\citep[e.g.][and references therein]{yakovlev2004}.
The cooling rate is sensitive to the physical properties of the superdense matter inside NSs, which are still poorly known \citep[e.g.][]{yakovlev2005}.
The equation of state (EoS) of such matter can be constrained by comparison of cooling theories with NSs surface temperatures derived from observational data.

The middle-aged PSR \jpsr\ (hereafter \psr) is a radio-quiet \gr\ pulsar 
discovered with the \fermi\ Large Area Telescope (LAT) \citep{SazParkinson2010}. 
Its spin period $P=375$ ms and period derivative $\dot{P}=7.1\times10^{-15}$ s~s$^{-1}$ imply the characteristic age 
$\tc\equiv P/2\dot{P}\approx 840$ kyr,
the spin-down luminosity $\dot{E}=5.3\times10^{33}$ \ergs\ 
and the characteristic (spin-down) magnetic field $B=1.65\times10^{12}$ G\footnote{Parameters are calculated using the timing solution for \psr\ based on five years of the \fermi\ data obtained from \url{https://confluence.slac.stanford.edu/display/GLAMCOG/LAT+Gamma-ray+Pulsar+Timing+Models}. See also \citet{kerr2015} for details.}.
The distance to the pulsar is poorly known. 
The only available estimate, 0.8 kpc, is the so-called `pseudo-distance' 
obtained using empirical correlation between the distance and the \gr\ flux 
above 100 MeV \citep{abdo2013}.
It is known to be uncertain by a factor of 2--3.
Analysing the $\gamma$-ray pulse profile, \citet{Pierbattista_15} estimated the magnetic inclination and line of sight angles of J1957  for different $\gamma$-ray emission geometries.

\begin{figure*}
\begin{minipage}[h]{0.495\linewidth}
\center{\includegraphics[width=0.95\linewidth,clip]{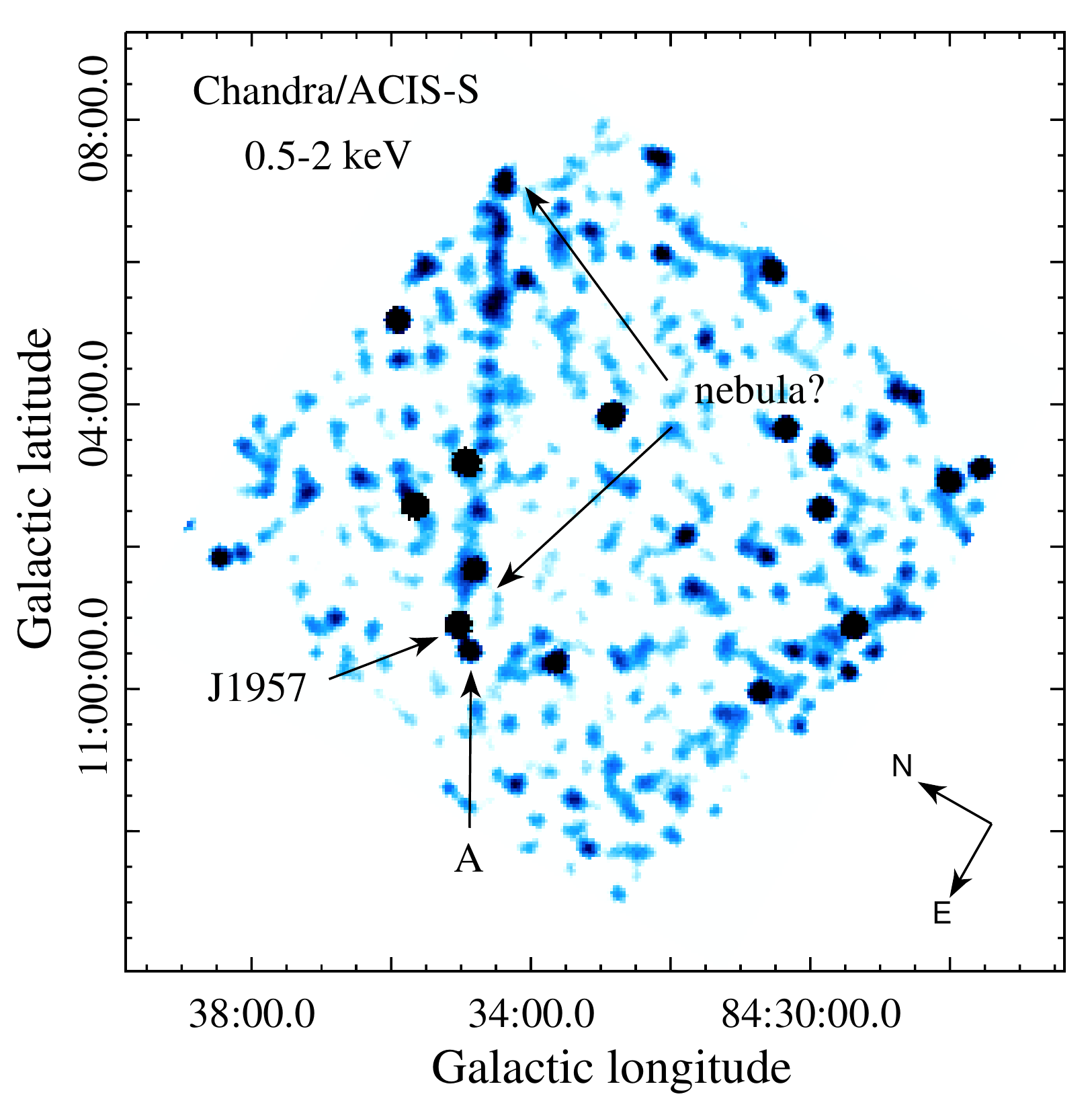}}
\end{minipage}
\begin{minipage}[h]{0.495\linewidth}
\center{\includegraphics[width=0.99\linewidth,clip]{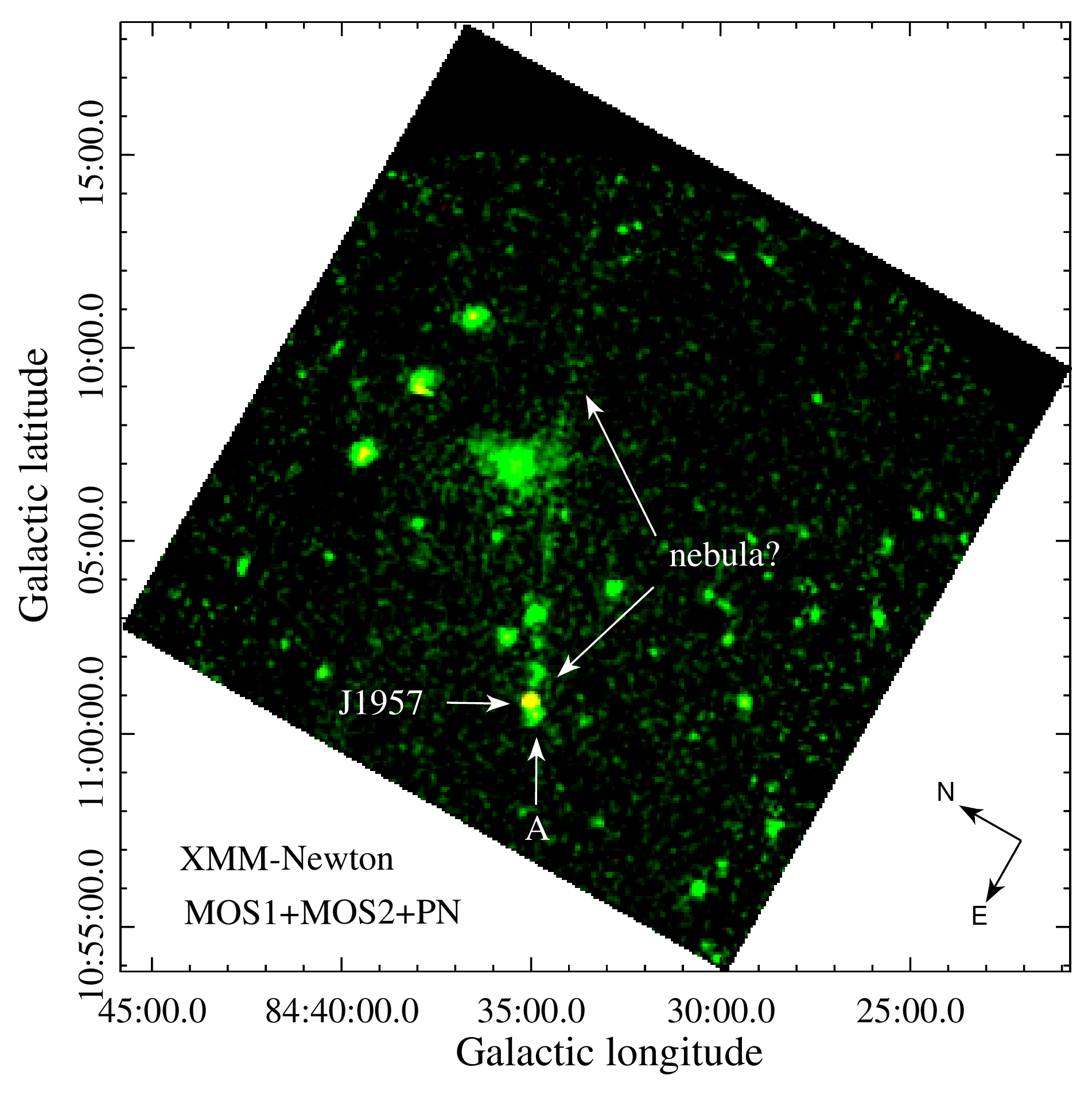}}
\end{minipage}
\caption{Images of the \psr\ field. 
\textit{Left:} \chan\ ACIS-S3 chip field-of-view in the 0.5--2 keV band.
\textit{Right:} 18 arcmin $\times$ 18 arcmin \xmm\ combined MOS1+MOS2+PN image
(red = 0.2--0.5 keV, green = 0.5--2 keV).
\psr, the nearby star `A' and the presumed trail-like nebula are marked. 
Compasses correspond to the equatorial coordinate system.
}
\label{fig:images}
\end{figure*}

The pulsar X-ray counterpart was identified by position in the 25-ks \chan\ Advanced CCD Imaging Spectrometer (ACIS-S) 
observation\footnote{ObsID 14828, PI M. Marelli, observation date 2014-02-01.} \citep{marelli2015}.
Its spectrum in the 0.3--10 keV band was found to be well described by the absorbed single power-law (PL)
with a photon index $\Gamma=2.1\pm0.3$, an energy integrated unabsorbed flux 
$F_X=(3.0\pm0.5)\times10^{-14}$ \flux\ 
and an absorption column density \nh\ $<2.5\times10^{20}$ cm$^{-2}$ 
\citep{marelli2015}.

 We reanalyzed the \chan\ data and confirmed the results by \citet{marelli2015}.
However, we found an unexpectedly large count number in the 0.1--0.3 keV band --
30 against 90 counts detected in the 0.3--10 keV band. 
This indicates 
the presence of a second soft component in the pulsar spectrum likely related 
to the thermal emission from the surface of the NS with a very low effective temperature.  
In this case, \psr\ becomes  especially  interesting for comparison with NS cooling theories according to which 
at  its characteristic age the pulsar should have already passed from a relatively slow neutrino cooling stage 
to a significantly faster photon stage where observational data on thermal emission from 
cooling NSs are particularly scarce.  
Unfortunately, the soft component cannot be confirmed using the \chan\ data 
since the ACIS energy scale is not calibrated below 0.3 keV. 3.2~s time resolution of the observations  
does not also allow to detect pulsations with the pulsar period.  
Therefore, we performed dedicated \xmm\ observations of \psr.
Here we present the analysis of these data. 
The X-ray data are described in Section~\ref{sec:data}. 
Timing and spectral analysis of \psr\ are presented in Sections~\ref{sec:timing} and \ref{sec:spectra}.
The results are discussed in Section~\ref{sec:discussion} and summarized in Section~\ref{sec:summary}.
Some details of the analysis are given in the Appendices.


\begin{figure*}
\begin{minipage}[h]{0.495\linewidth}
\center{\includegraphics[width=0.93\linewidth,clip]{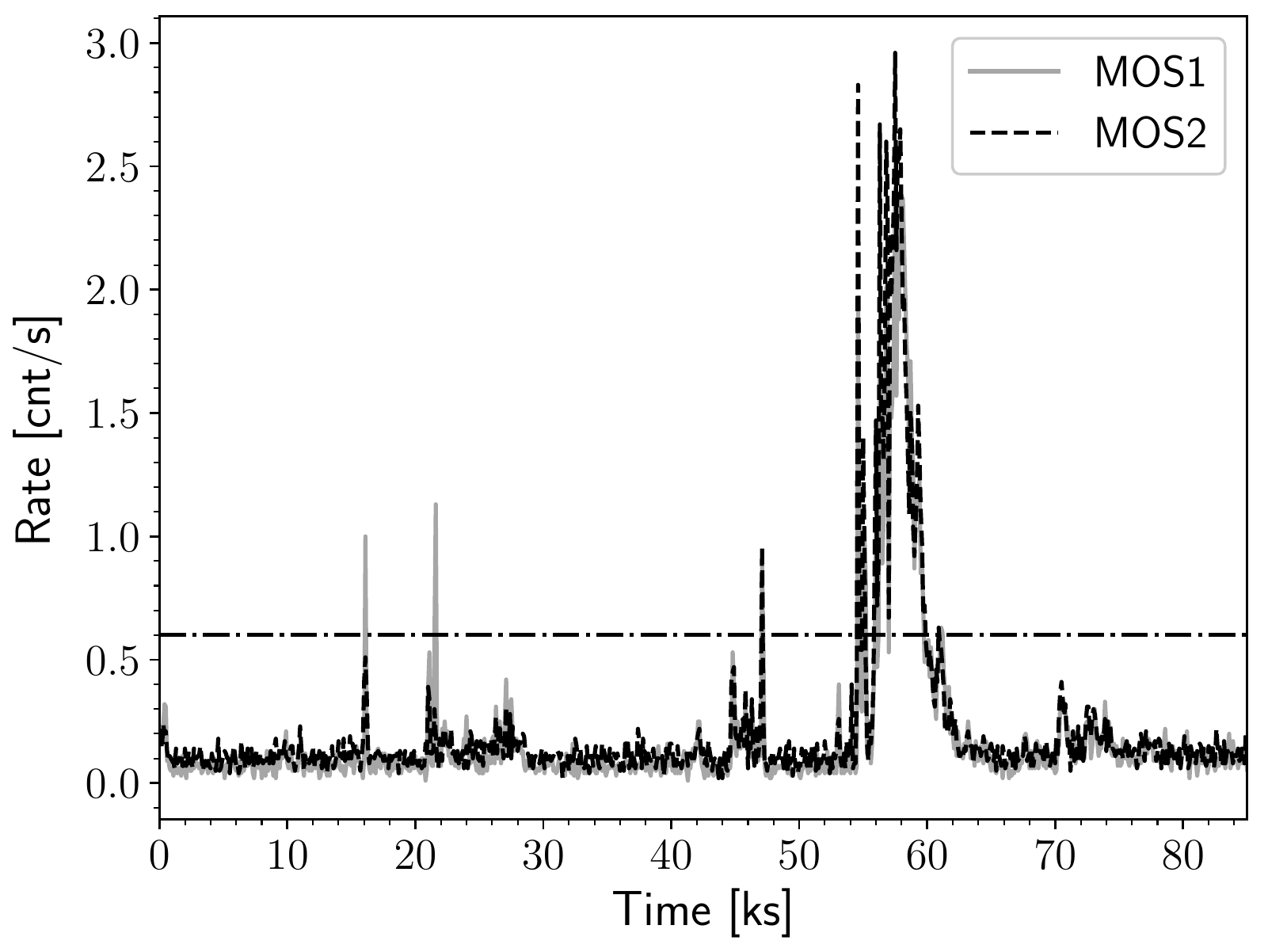}}
\end{minipage}
\begin{minipage}[h]{0.495\linewidth}
\center{\includegraphics[width=0.93\linewidth,clip]{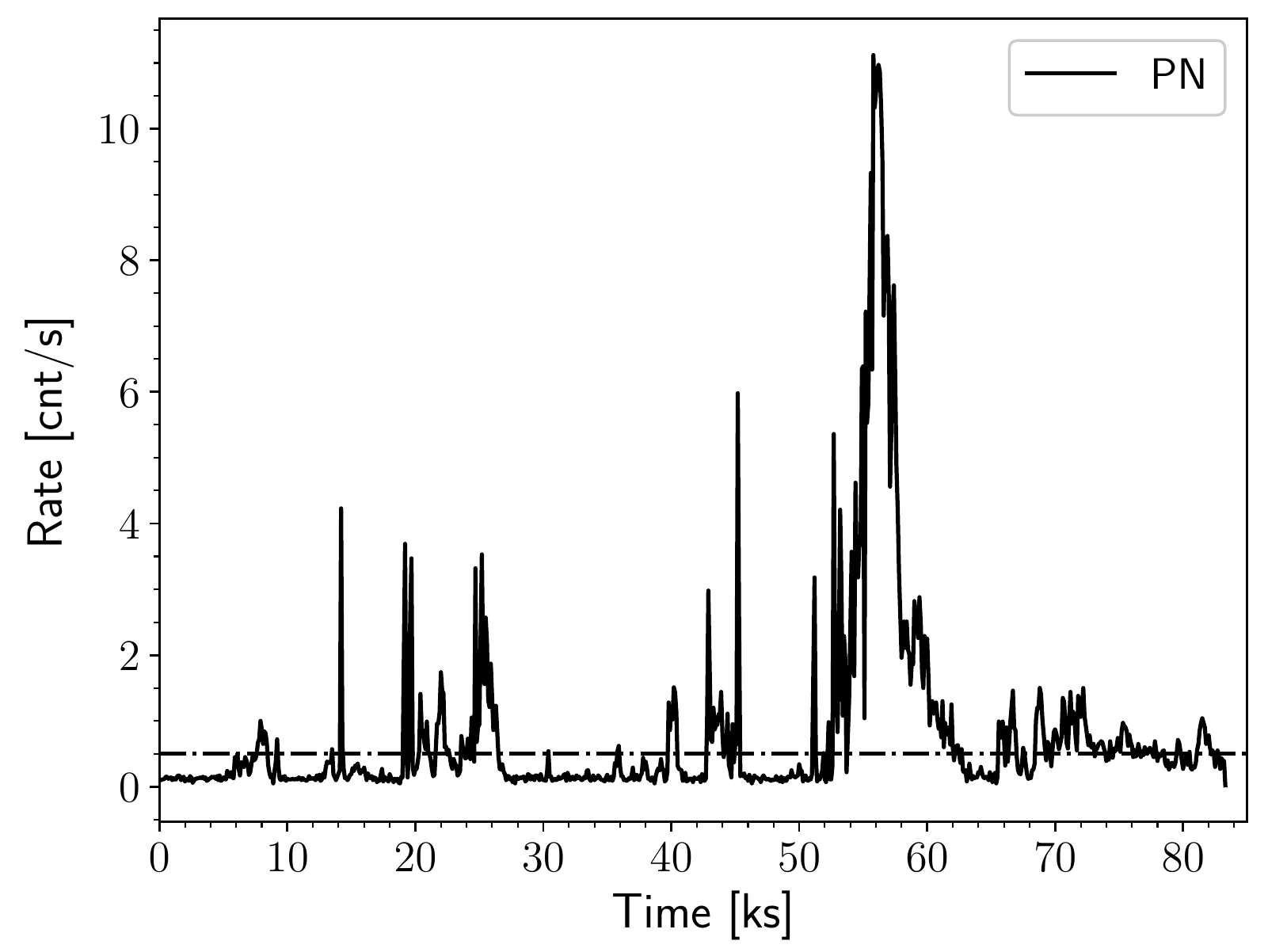}}
\end{minipage}
\caption{High-energy light curves obtained from the FOVs of MOS (10--12 keV; \textit{left}) and pn (12--14 keV; \textit{right}) detectors. Dash-dotted lines indicate thresholds used to filter out periods of background flares.}
\label{fig:rate}
\end{figure*}

\section{X-ray data and imaging}
\label{sec:data} 

The \psr\ field was observed with \xmm\ on 2019 October 5 (ObsID 0844930101, PI D. Zyuzin).
The total exposure was about 87 ks.
The European Photon Imaging Camera Metal Oxide Semiconductor (EPIC-MOS) detectors were operated in the full-frame mode with the imaging area of about 28 arcmin $\times$ 28 arcmin and the medium filter and the EPIC-pn camera -- in the large window mode with the imaging area of about 13.5 arcmin $\times$ 26 arcmin and the thin filter.
We also used the \chan/ACIS-S archival dataset (ObsID 14828) where the pulsar was exposed on the S3 chip.
To analyze the data, we utilized the \xmm\ Science Analysis Software (\textsc{xmm-sas}) v.~17.0.0 and \chan\ Interactive Analysis of Observations (\textsc{ciao}) v.~4.12 packages.

The \chan\ dataset was reprocessed using the \textsc{chandra\_repro} tool. 
Applying the \textsc{fluximage} task, we created the exposure-corrected image of the ACIS-S3 chip which is presented in the left panel of Fig.~\ref{fig:images} where the pulsar counterpart  and the nearby star `A' are marked.
The \textsc{wavdetect} command was used to obtain coordinates of point-like sources.
For \psr, we derived R.A.~=~19\h57\m38\fss390(6) and Dec.~=~+50\degs33\amin21\farcs02(5) 
(numbers in parentheses are 1$\sigma$ pure statistical uncertainties).

We combined the data from both MOS and PN detectors to obtain deeper X-ray images using 
the `images' script\footnote{\url{https://www.cosmos.esa.int/web/xmm-newton/images}.} \citep{imagesscript}.
The resulting image 
is shown in the right panel of Fig.~\ref{fig:images}.
Since \xmm\ has lower spatial resolution than \chan, \psr\ is somewhat blurred with the star `A'. 
One can see a faint thin feature protruding  
from \psr\ almost perpendicularly to the Galactic plane. 
It has a clumpy structure and is also visible in the \chan\ image where it extends up to the edge of the ACIS-S detector ($\approx 6$ arcmin). 
In the \xmm\ image its length seems to be longer, at least up to $\sim$8 arcmin, though its faintness and some blurring with other sources 
preclude accurate measurements.
This maybe a trail-like pulsar wind nebula (PWN) powered by \psr. 
One  can  see that the pulsar is indeed a rather soft source while the presumed PWN is produced by harder photons.

For the further analysis, we filtered out the background flares 
inspecting high-energy light curves extracted from the field-of-views (FoVs) of all EPIC detectors.
We chose the following threshold count rates to define good time intervals: 0.5 counts~s$^{-1}$ for pn 
and 0.6 counts~s$^{-1}$ for both MOS cameras (see Fig.~\ref{fig:rate}).
The resulting effective exposures are about 79.8, 79.8 and 48.7 ks for the MOS1, MOS2 and pn detectors, respectively.
We selected single to quadruple pixel events (\textsc{pattern} $\leq12$) for the MOS data and single and double pixel events (\textsc{pattern} $\leq4$) for the pn data.


\section{Timing}
\label{sec:timing}

\begin{figure}
\begin{minipage}[h]{1.\linewidth}
\center{\includegraphics[width=1.\linewidth,clip]{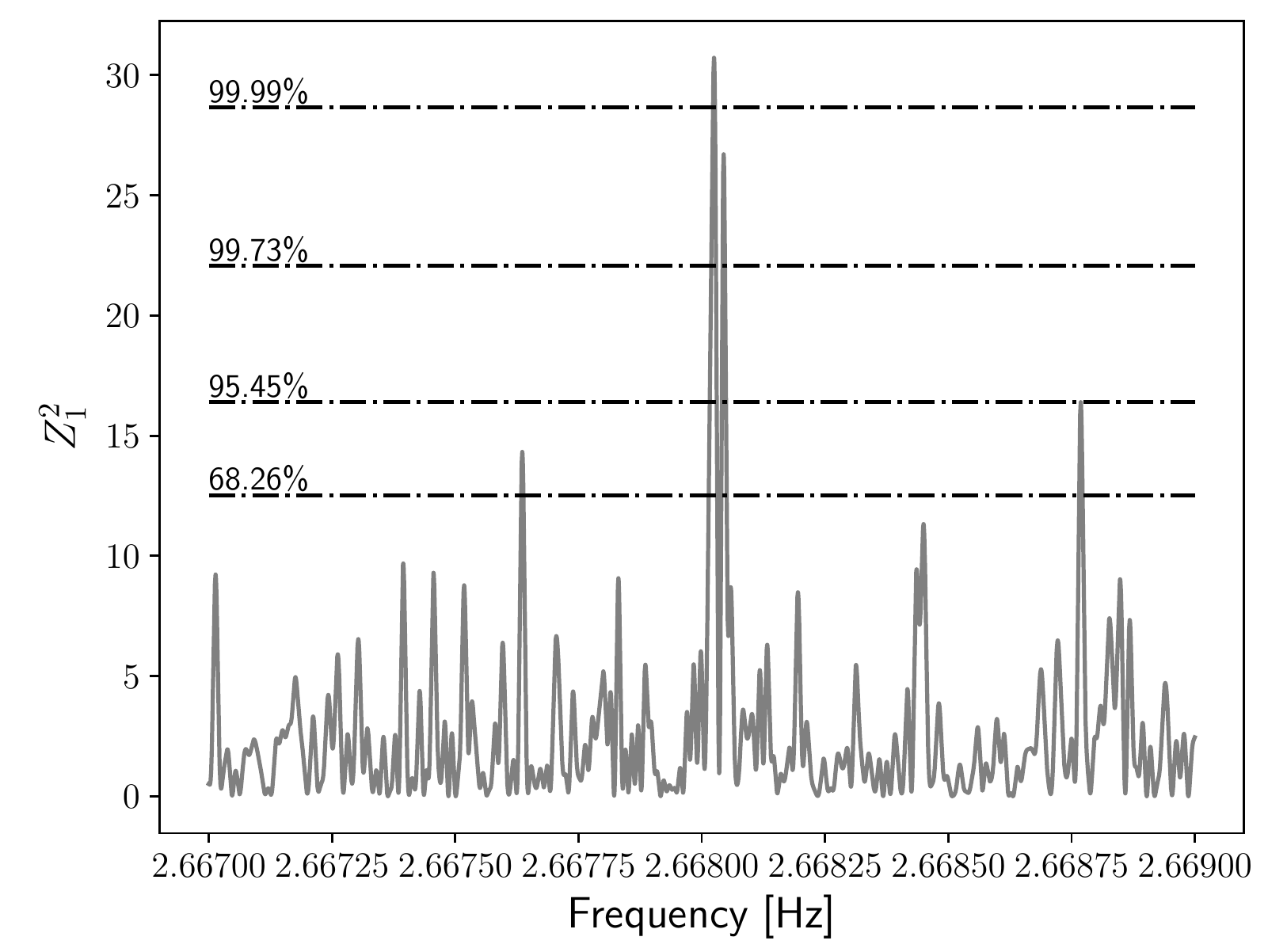}}
\end{minipage}
\caption{$Z^2_1$-test periodogram for \psr. Dashed lines show confidence levels.}
\label{fig:ztest}
\end{figure}

\begin{figure}
\begin{minipage}[h]{1.\linewidth}
\center{\includegraphics[width=0.98\linewidth,clip]{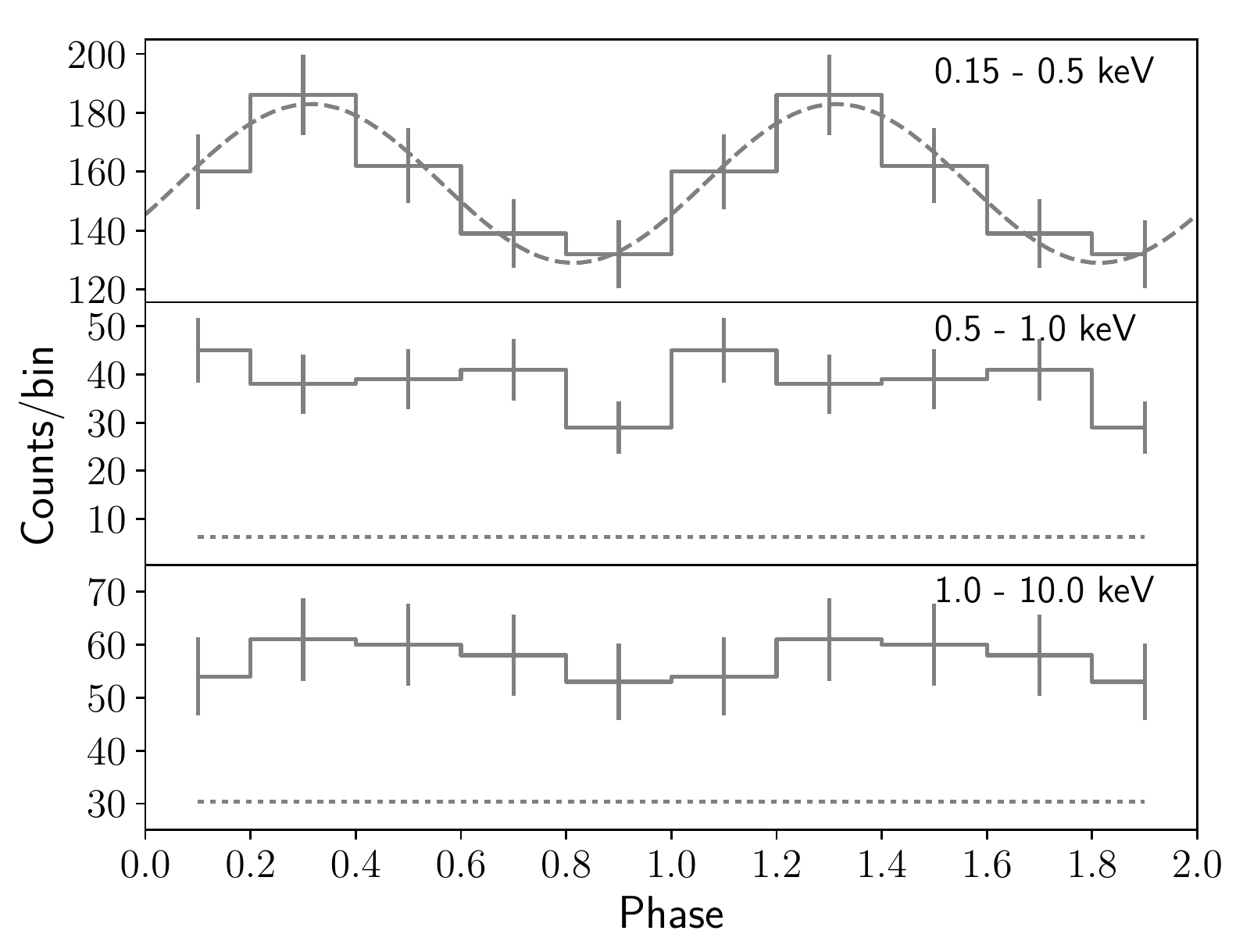}}
\end{minipage}
\caption{Folded X-ray light curves for \psr\ in different energy bands indicated in the panels (periods of background flares are removed). Dotted lines in the middle and bottom panels indicate the background level. In the 0.15--0.5 keV band the background contribution ($\approx 12$ counts phase bin$^{-1}$) is very low in comparison with the pulsar count rate and thus it is not shown. The best-fitting sine curve is overlaid.}
\label{fig:prof}
\end{figure}

The timing resolution of the EPIC-pn detector operating in the large window mode is $\approx48$ ms 
which allows us to search for pulsations from \psr.
We used the event list unfiltered from background flares since multiple time gaps in the data may hamper signal detection. 
The barycenter correction was applied by the \textsc{xmm-sas barycen} command using DE405 ephemeris 
and the pulsar coordinates derived from the \chan\ data.
We extracted events in the 0.15--0.5 keV band using the 12.5-arcsec radius circle around the \chan\ position of \psr. Such a small aperture was chosen to eliminate the contribution from the star `A' located at about 20 arcsec from the pulsar.  We searched for pulsations 
utilising $Z^2_1$-test \citep{ztest} and 2.667--2.669 Hz frequency range,
encapsulating the predicted pulsar rotation frequency 
of about 2.6680281 
Hz 
obtained from 
extrapolation of the \fermi\ timing solution to the epoch of the \xmm\ observations (MJD 58761), and
with a step of 0.1 $\mu$Hz.

The resulting periodogram is shown in Fig.~\ref{fig:ztest}.
The maximum $Z^2_{1}$ is $\approx$ 30.7. 
This implies the confidence level of a detection $C=[1-\mathcal{N}\mathrm{exp(-Z^2_{1,max}/2)}]\times100\% =99.996$ per cent 
(or $\approx 4\sigma$) where $\mathcal{N}=\Delta f\times T_\mathrm{obs}$ is the number of statistically independent trials, 
$\Delta f$ is the frequency range and $T_\mathrm{obs}$ is the duration of the observation.
The corresponding frequency is 2.6680249(12) Hz (the frequency 1$\sigma$ uncertainty was calculated using the formula from \citealt{chang2012}). 
This is consistent within 3$\sigma$ with the predicted value 
from 
\fermi\ timing solution 
and firmly establishes the pulsar nature of the X-ray source.

The \psr\ X-ray pulse profile obtained using the derived frequency is presented in Fig.~\ref{fig:prof}. 
We see that pulsations are clearly detected only in the very soft band.
Non-detection in harder bands is likely due to the low number of counts from the pulsar. 
The background-corrected pulsed fraction in the 0.15--0.5 keV band $\mathrm{PF}=(f_\mathrm{max}-f_\mathrm{min})/(f_\mathrm{max}+f_\mathrm{min})$, where $f_\mathrm{max}$ and $f_\mathrm{min}$ are the maximum and minimum intensity of the folded light curve, is $\approx18\pm6$ per cent. 
As an additional check, 
we fitted the pulse profile in the  0.15--0.5 keV band with a sine function (fundamental component; see Fig.~\ref{fig:prof})  and got the same result. 
Following the method from \citet{brazier1994}, we estimated the 99 per cent upper limits on the pulsed fraction of $\approx44$ per cent in the 0.5--1 keV band and $\approx64$ per cent in the 1--10 keV band.



\section{Spectral analysis}
\label{sec:spectra}

\begin{figure}
\begin{minipage}[h]{1\linewidth}
\center{\includegraphics[width=0.99    \linewidth,clip]{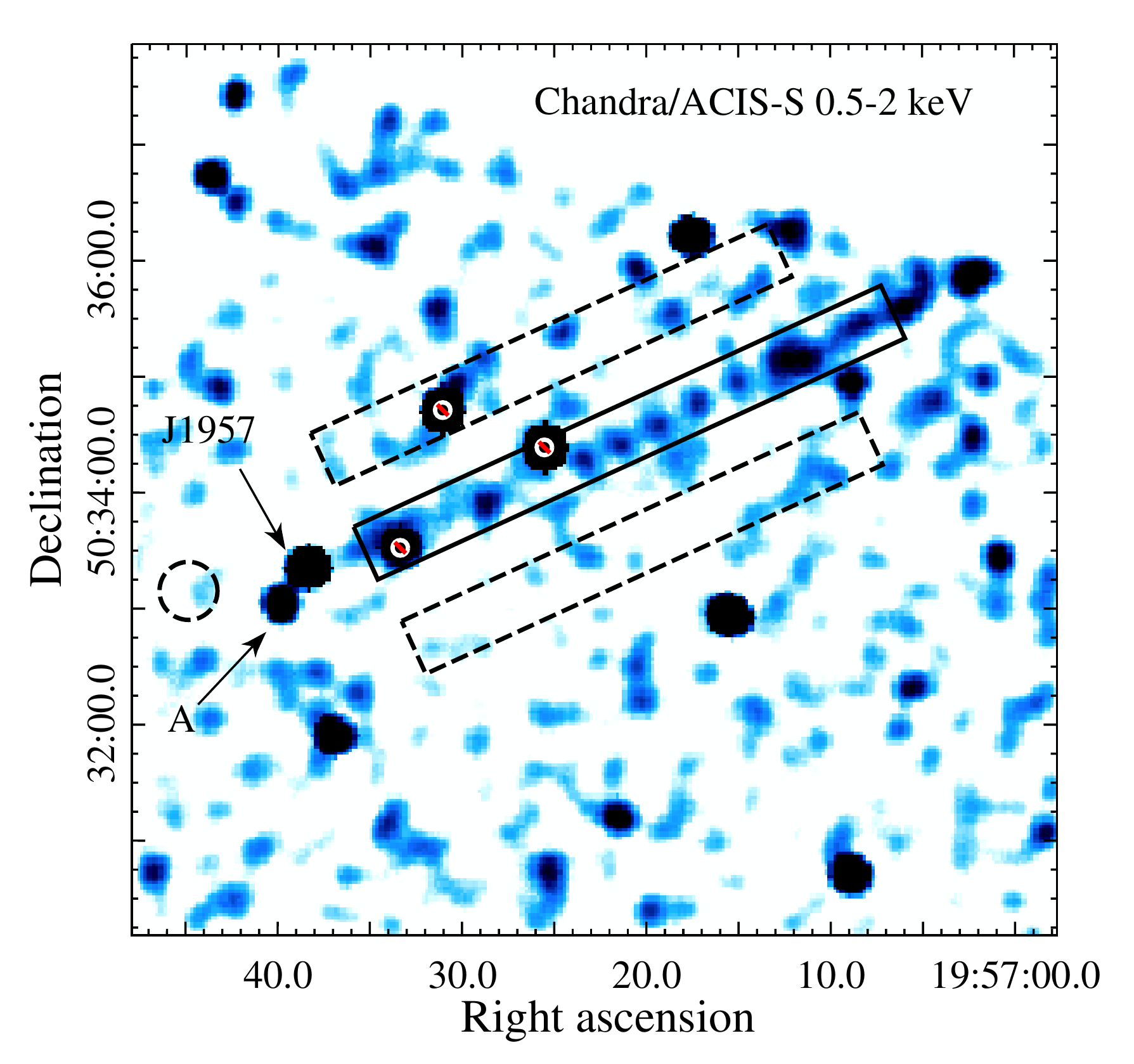}}
\end{minipage}
\caption{\chan\ ACIS image of the \psr\ field in the 0.5--2 keV band. \psr\ and the nearby star `A' also marked.
The background for the pulsar was extracted from the dashed circle. 
The solid rectangle shows the region \textbf{used} to extract the spectrum of the trail-like nebula
while the dashed rectangles were used for the background.
Stars excluded from the regions are shown by crossed circles.
}
\label{fig:tail_reg}
\end{figure}

\subsection{\psr}
\label{subsec:psr_spec}

\begin{figure*}
\begin{minipage}[h]{1.\linewidth}
\center{\includegraphics[width=0.96\linewidth,clip]{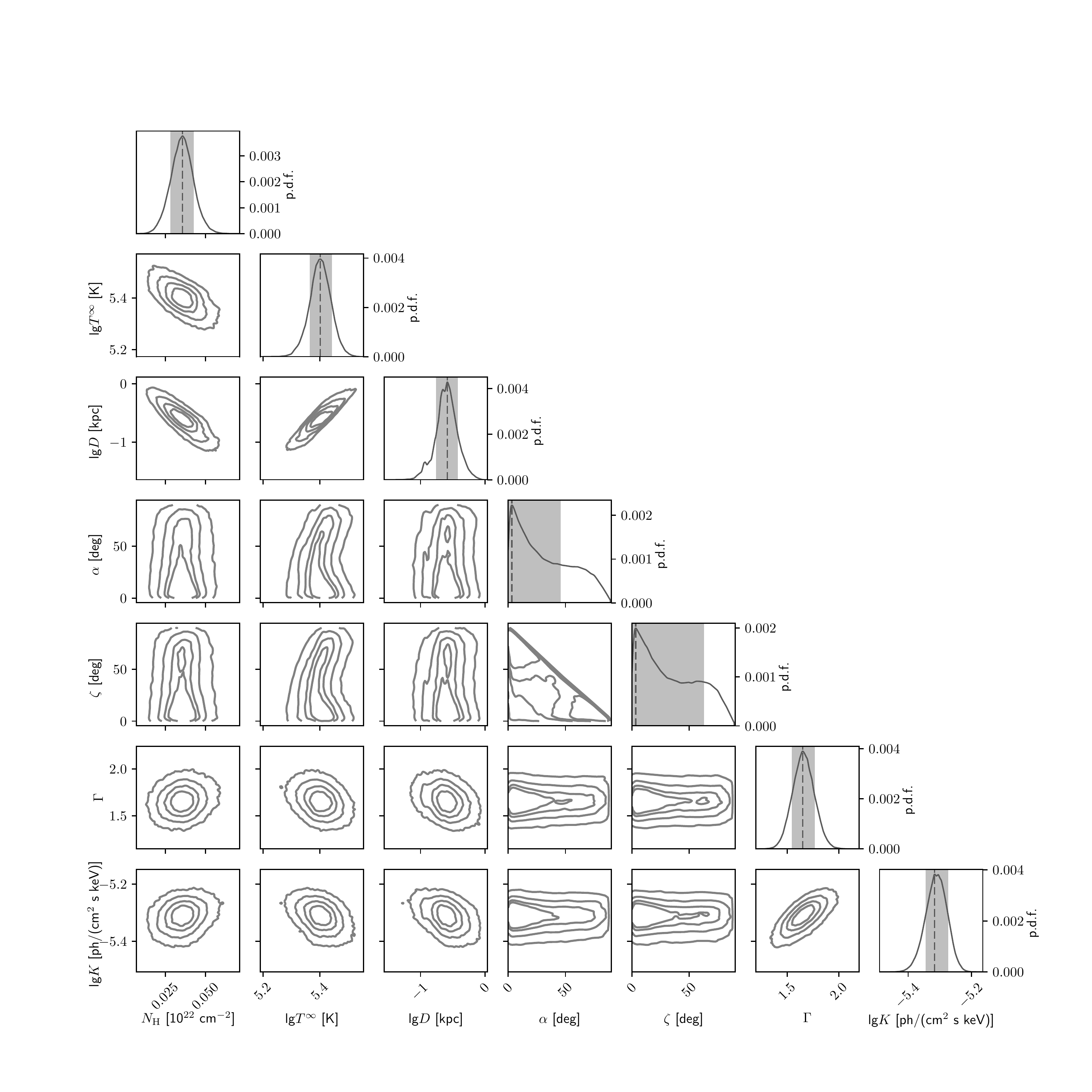}}
\end{minipage}
\caption{1D and 2D marginal posterior distribution functions (p.d.f.) for parameters of the \textsc{nsmdip1}$+$PL model (see Table~\ref{tab:psrpar}). 
Vertical dashed lines in 1D distributions indicate the best-fitting values while light gray strips show 1$\sigma$ credible intervals. 
In 2D distributions, 40, 68, 90 and 99 per cent confidence contours are shown.
}
\label{fig:triang}
\end{figure*}

We extracted the time integrated pulsar spectra from both  the \xmm\ and \chan\ data.
In the latter case we utilized the \textsc{specextract} routine and the 2.5-arcsec radius aperture.
The \xmm\ spectra were extracted from the 12.5-arcsec radius circle using \textsc{evselect} task 
and redistribution matrix and ancillary response files were created by \textsc{rmfgen} and \textsc{arfgen} commands. 
The background spectrum was obtained from the source-free region (see Fig.~\ref{fig:tail_reg}).
For the interstellar medium (ISM) absorption, we applied \textsc{tbabs} model with the \textsc{wilm} abundances \citep*{wilms2000}.
The spectra were 
fitted simultaneously in the X-Ray Spectral Fitting Package \textsc{xspec} 
v.12.10.1\footnote{\url{https://heasarc.gsfc.nasa.gov/docs/xanadu/xspec/}} \citep*{xspec}.
We used the following energy ranges: 0.3--10 keV for the \chan, 0.2--10 keV for the MOS and 0.15--10 keV for the pn spectra.
The resulting number of source counts after background subtraction is 
232(MOS1) + 254(MOS2) + 902(pn) + 88(ACIS).

As a first step, to check how  different models fit the data, we applied the  $\chi^2$-statistic and grouped the
data to ensure 25 counts per energy bin.
The single absorbed PL (which describes the pulsar non-thermal emission of magnetospheric origin) 
or blackbody (BB, which describes the thermal emission from the NS surface) models 
resulted in unacceptable fits with reduced $\chi^2_\nu$ = 2.36
and 5.5 for 54 degrees of freedom (dof), respectively.
Then we tried the composite BB + PL model
and found that it fits the spectra well with $\chi^2_\nu$ = 1.13 (52 dof).

For the thermal component, we also tried the NS magnetized atmosphere models \textsc{nsmaxg} \citep*{ho2008} 
presented in the \textsc{xspec} package assuming an NS mass $M_\mathrm{NS}$ = 1.4\msun\ and radius $R = 13$ km.
However, in this case the temperature tends to the lowest value available for these models 
$\mathbf{\lg(T/{\rm K})=5.5}$; here and hereafter $\lg\equiv\log_{10}$. 
Therefore, we calculated another grid of NS atmosphere models, 
\textsc{nsmdip},  which include lower temperatures for the magnetic fields
that seem to be likely for this pulsar. We assumed a dipole magnetic field
(taking the effects of General Relativity into account)
and a corresponding distribution of the local effective temperature over the stellar surface. 
In these models, 
the angle $\alpha$ between the rotation and the magnetic axes and
the angle $\zeta$ between the rotation axis and the line of sight
can be used as  free parameters. 
The total thermal luminosity $L^\infty$
that would be measured by a distant observer is calculated
by a proper surface integration and converted into 
the global effective temperature $T^\infty$.
The details of calculations are presented in Appendix \ref{appendix}.

We have tested several \textsc{nsmdip} models (Table~\ref{tab:dipset}). In model \textsc{nsmdip1}, 
we have assumed the canonical NS mass $M=1.4\msun$ and 
magnetic field at the pole $B_\mathrm{p}=3\times10^{12}$~G,
which corresponds to the characteristic magnetic field
$B\approx1.65\times10^{12}$~G of \psr\ at the equator derived from the spin-down. The radius $R=12.6$ km was taken 
according to the EoS BSk24 \citep{Pearson_18};
the corresponding gravitational redshift is $z_\mathrm{g}=0.22$, 
and surface gravity $g_\mathrm{s}=1.43\times10^{14}$ cm s$^{-2}$.
In order to test a higher redshift, we consider a more compact NS model (\textsc{nsmdip2})  
with $M=2\,\msun$ and $R=11.4$ km
($z_\mathrm{g}=0.44$ and
$g_\mathrm{s}=3\times10^{14}$ cm s$^{-2}$),
which approximately corresponds to the EoS BSk26.
Model \textsc{nsmdip3}  has the same $z_\mathrm{g}$ and $R$ as model 1
but a more consistent magnetic field estimate. 
The canonical characteristic magnetic field 
is the equatorial field of an orthogonal
rotating dipole in vacuum with
the given spin period and its derivative,
assuming $R=10$ km
and $I_{45}=1$, where $I_{45}$
is the moment of inertia in units of $10^{45}$ g cm$^2$
(e.g., \citealt{ManchesterTaylor}). 
However, according to the EoS BSk24, 
for $M=1.4\,\msun$ we have $R=12.6$ km and $I_{45}=1.51$.
Moreover, the spin-down of a pulsar is affected by its magnetosphere.
Results of numerical simulations of plasma
behaviour in the pulsar magnetosphere suggest that 
the characteristic magnetic field should be multiplied by  
a factor of $0.8\,
(R/\mbox{10 km})^{-3}\,(1+\sin^2\alpha)^{-1/2}\,\sqrt{I_{45}}$,
where $\alpha$ is the angle between rotational and magnetic axes
\citep{Spitkovsky06}. For the above-mentioned values of $R$ and $I_{45}$ this implies
a 2--3 times weaker field compared to
the characteristic one. Thus we adopted
$B_\mathrm{p}=1.1\times10^{12}$~G in model \textsc{nsmdip3}
which corresponds to the field strength at the equator
$B_\mathrm{eq} \approx 6\times10^{11}$~G, that is 2.7 times smaller than the pulsar characteristic (spin-down) field.
We find that all three absorbed \textsc{nsmdip}$+$PL models 
describe the data equally well as the BB$+$PL model giving $\chi^2_\nu\approx$ 1.17 (51 dof). 

Since the number of source counts is not large, in order to get the most robust estimates of 
the model parameters and their uncertainties from spectral fits, 
we regrouped all spectra to ensure at least 1 count per energy bin and therefore
used $W$-statistic \citep{xspec} which is $C$-statistic \citep{1979cash} suitable for Poisson data with Poisson background.
We then performed the fitting using a Markov chain Monte-Carlo
(MCMC) sampling procedure. 
We employed the affine-invariant MCMC sampler developed by \citet{goodman&weare2010} and 
implemented in a \textsc{python} package \textsc{emcee} by \citet{emcee2013}. 
In addition, to estimate the distance to \psr, we used the interstellar absorption--distance relation 
 towards the pulsar as a prior (see Appendix~\ref{appendix:nh-d} for details). 
About 100 walkers and 13000 steps were typically enough to ensure fit convergences.
Using the sampled posterior distribution, 
we obtained the best-fitting parameters of the models with uncertainties, which are defined as their maximal-probability density values and 
respective credible intervals. 

\begin{table*}
\renewcommand{\arraystretch}{1.2}
\caption{Best-fitting parameters for different models with thawed  
$\alpha$ and $\zeta$ angles and accounting for $\alpha$ + $\zeta$ <= 90$\degs$.
}
\label{tab:psrpar}
\begin{center}
\begin{tabular}{lccccccccc}                         
\hline 
Model & \nh,                &  $k_\mathrm{B} T^\infty$, & $R^\infty$, & $\lg\,L^\infty$, & $\Gamma$  & $K$,  & $D$, & $-\ln \mathcal{L}$ & BIC\\
      & 10$^{20}$ cm$^{-2}$ & eV                     & km          & \ergs            &           & \phs  & pc   &               &    \\
\hline
BB + PL & $2.2^{+1.6}_{-1.0}$ & $54^{+5}_{-7}$ & $2.2^{+5.0}_{-2.0}$ & $30.66^{+0.85}_{-1.72}$  & $1.76^{+0.11}_{-0.11}$ & $5.4^{+0.4}_{-0.4}\times 10^{-6}$ & $545^{+463}_{-445}$ &  193.5 & 416.6 \\ 

\hline
\textsc{nsmdip1}  + PL & $3.6^{+0.6}_{-0.5}$ & $21.7^{+2.0}_{-1.7}$ & 15.4$^f$ & $30.83^{+0.16}_{-0.14}$ & $1.65^{+0.11}_{-0.10}$ & $4.8^{+0.5}_{-0.3}\times 10^{-6}$ & $260^{+109}_{-82}$  & 195.3 & 432 \\
       
\textsc{nsmdip2}  + PL & $3.2^{+0.6}_{-0.7}$ & $19.8^{+2.0}_{-1.8}$ & 16.4$^f$ & $30.72^{+0.17}_{-0.16}$ & $1.66^{+0.10}_{-0.11}$ & $4.9^{+0.4}_{-0.4}\times 10^{-6}$ & $218^{+164}_{-64}$  & 195.9 & 433\\
        
\textsc{nsmdip3} + PL & $2.6^{+0.8}_{-0.5}$ & $18.4^{+2.1}_{-2.0}$ & 15.4$^f$ & $30.55^{+0.19}_{-0.21}$ & $1.64^{+0.10}_{-0.10}$ & $4.9^{+0.3}_{-0.4}\times 10^{-6}$ &  $210^{+108}_{-106}$ & 196.1 & 433.7 \\ 
     
\hline
\end{tabular}
\end{center}
\begin{tablenotes}
\item $^\dag$ \nh\ is the absorbing column density, 
$T^\infty = T/(1+z_\mathrm{g})$ is the effective temperature as measured by a distant observer, 
$R^\infty=R(1+z_\mathrm{g})$ is the radius of the equivalent emitting sphere as seen by a distant observer,
$L^\infty = L/(1+z_\mathrm{g})^2=4\pi R^2\sigma_{\rm SB}T^4/(1+z_\mathrm{g})^2$ is the bolometric thermal 
luminosity as measured by a distant observer ($\sigma_{\rm SB}$ is the Stefan-Boltzmann constant), 
$\Gamma$ is the photon index, $K$ is the PL normalization and $D$ is the distance.
For the gravitational redshift $z_\mathrm{g}$ and unredshifted radius $R$, see Table~\ref{tab:dipset}. The last two columns give the values of
the maximum log-likelihood $\ln \mathcal{L}$ and
the Bayesian information criterion (BIC).
All errors 
are at 1$\sigma$ credible interval.
\item $^f$ Fixed parameters.
\end{tablenotes}
\end{table*}
\begin{table*}
\renewcommand{\arraystretch}{1.2}
\caption{
The same as in Table~\ref{tab:psrpar} but for fixed angles at
$\alpha$ = $66\degs$ and  $\zeta$ = $24\degs$ $^\dag$.  
}
\label{tab:psrpar_fixed_angle}
\begin{center}
\begin{tabular}{lccccccccc}                         
\hline 
Model & \nh,  & $k_\mathrm{B} T^\infty$, & $R^\infty$, & $\lg\,L^\infty$, & $\Gamma$  & $K$,  & $D$, & $-\ln \mathcal{L}$ & BIC  \\  
      & 10$^{20}$ cm$^{-2}$ & eV                       & km          & erg s$^{-1}$     &           & \phs  & pc   &               &       \\
\hline
\textsc{nsmdip1}  + PL & $3.5^{+0.7}_{-0.6}$ & $21.9^{+1.9}_{-1.1}$ & 15.4$^f$ & $30.85^{+0.14}_{-0.09}$ & $1.66^{+0.10}_{-0.11}$ & $4.9^{+0.4}_{-0.4}\times 10^{-6}$ & $263^{+112}_{-78}$  & 195.1  & 419.8 \\

\textsc{nsmdip2}  + PL & $3.3^{+0.6}_{-0.8}$ & $20.3^{+2.0}_{-1.5}$ & 16.4$^f$ & $30.77^{+0.16}_{-0.14}$ & $1.64^{+0.11}_{-0.09}$ & $5.0^{+0.3}_{-0.5}\times 10^{-6}$ & $224^{+165}_{-63}$  & 195.7  & 421.0\\

\textsc{nsmdip3} + PL & $2.9^{+0.5}_{-0.7}$ & $19.8^{+1.5}_{-2.6}$ & 15.4$^f$ & $30.67^{+0.13}_{-0.24}$ & $1.63^{+0.11}_{-0.10}$ & $4.8^{+0.3}_{-0.4}\times 10^{-6}$ & $212^{+99}_{-110}$  & 195.9 & 422.2 \\
\hline
\end{tabular}
\end{center}
\begin{tablenotes}
\item $^\dag$ Notations are the same as in Table~\ref{tab:psrpar}. 
\end{tablenotes}
\end{table*}

\begin{figure}
\begin{minipage}[h]{1.\linewidth}
\center{\includegraphics[width=0.99\linewidth,clip]{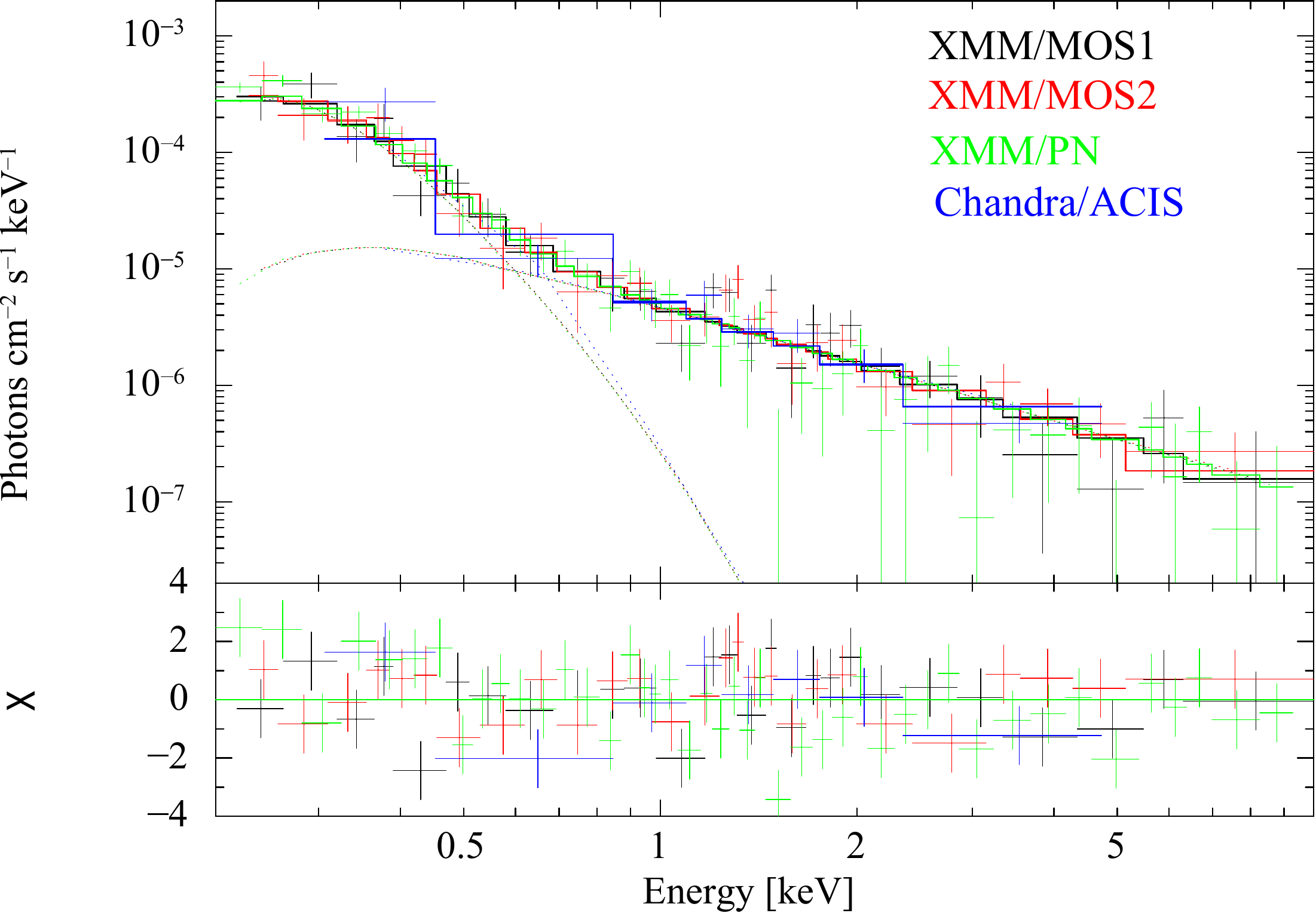}}
\end{minipage}
\caption{The \psr\ unfolded spectrum and the model \textsc{nsmdip1}$+$PL with 
fixed angles at
$\alpha$ = $66\degs$ and  $\zeta$ = $24\degs$
(see the first line in Table~\ref{tab:psrpar_fixed_angle}). 
Dotted lines show the model components. 
Data from different instruments are indicated by different colours as indicated in the panel.
Spectra were grouped to ensure at least 10 counts per energy bin for illustrative purposes.}
\label{fig:spec}
\end{figure}

In the case of  \textsc{nsmdip}$+$PL models, we found that our data are rather insensitive 
to the $\alpha$ and $\zeta$ angles.
From Fig.~\ref{fig:prof}, one can see that there is one pulse per period.
This implies that  $\alpha+\zeta \leq 90$\degs, which was used as a prior in the spectral fitting.
As an example, 1D and 2D marginal posterior parameter distribution functions for 
the \textsc{nsmdip1}$+$PL model are shown in Fig.~\ref{fig:triang}. 
One can see that 
the angles, in contrast to other parameters,  cannot be well constrained from the fit.
Their formal best-fitting values are close to 
0\degs. 
Our fits show that 
the similar situation occurs for the 
\textsc{nsmdip2}$+$PL and \textsc{nsmdip3}$+$PL models.
However, for any of the models the probability density function for 
angles remains  relatively high for the whole meaningful ranges of 0\degs -- 90\degs.
This results in a very large formal uncertainties of the angles, 
and in fact they can accept any value. 
For these reasons, 
angles are not included in the parameter list 
of Table~\ref{tab:psrpar} presenting the fit results. 
We found that other fit parameters depend on a specific value of angles only weakly. 
We also fitted spectra using two models of the \psr\ $\gamma$-ray emission geometry, polar cap (PC) and slot-gap (SG),
obtained from the \fermi\ data by \citet{Pierbattista_15} which satisfy the condition $\alpha+\zeta \leq 90$\degs\footnote{
\citet{Pierbattista_15} used  polar cap (PC), slot gap (SG), outer gap (OG) and one pole caustic (OPC) models 
which assume different regions of the pulsar magnetosphere where particles are accelerated and emit $\gamma$-rays. 
In the PC model this occurs 
at low altitudes near the magnetic poles and in the SG model -- in a slot gap,  
which is a narrow gap extending from the polar cap surface to the light cylinder. 
In the OG model the gap extends from the null charge surface to high altitudes along the last-open-field lines. 
The OPC  is a variation of the OG model which suggests different gap width and energetics.  
The SG and OG  can provide wide \gr\ beams and imply exponential spectral cut-off at high energies 
while the PC  provides narrow beams and predicts super-exponential spectral cut-off  due to magnetic pair production process.}.
The results for the SG with $\alpha=66\degs$ and $\zeta=24\degs$ are presented in Table~\ref{tab:psrpar_fixed_angle} and Fig.~\ref{fig:spec}.
This model geometry is more preferable as it gives  an acceptable X-ray pulse shape and pulsed fraction (see Section~\ref{sec:discussion} for details).


To understand which of the models is statistically more  preferable, 
Tables~\ref{tab:psrpar} and \ref{tab:psrpar_fixed_angle} 
also present a Bayesian evidence $\mathcal{L}$ and information criteria 
BIC = $n\ln N_{\rm b} -2\ln \mathcal{L}$, where $n$ is the number of free model parameters 
and $N_{\rm b}$ is the number of spectral bins (which is 373). 
When picking from several models, the one with the lowest BIC is preferred. As seen, 
the BB$+$PL model has the smallest BIC and appears to be preferable.  
The strength of the evidence against \textsc{nsmdip}$+$PL models 
with the higher BICs is defined by $\Delta$BIC$\ga 15$ for atmosphere models with free angles 
which is evaluated as a \textit{very strong evidence}. 
At the same time, $\Delta$BICs for any pair of \textsc{nsmdip}$+$PL 
models in Table~\ref{tab:psrpar} is $\la 2$, which is qualified either 
as weakly \textit{positive} or \textit{not worth more than a bare mention}. 
On the other hand, fixing of $\alpha$ and $\zeta$ almost does not change  $\mathcal{L}$ 
while BICs values become smaller.
In this case, the difference between \textsc{nsmdip}$+$PL and BB+PL models $\Delta$BIC$ \la 6$ which changes the strength of evidence from  \textit{very strong} to \textit{decisive}.
Moreover, the models \textsc{nsmdip}+PL are preferred from the physics point of view,
because they assume the plausible NS radii and temperature distributions,
whereas the BB+PL model results in a best-fitting radius incompatible
with thermal emission from the entire NS surface (see the discussion in Section~\ref{sec:discussion}).

\subsection{The trail-like nebula}
\label{subsec:tail_spec}

\begin{figure}
\begin{minipage}[h]{1\linewidth}
\center{\includegraphics[width=0.99\linewidth,clip]{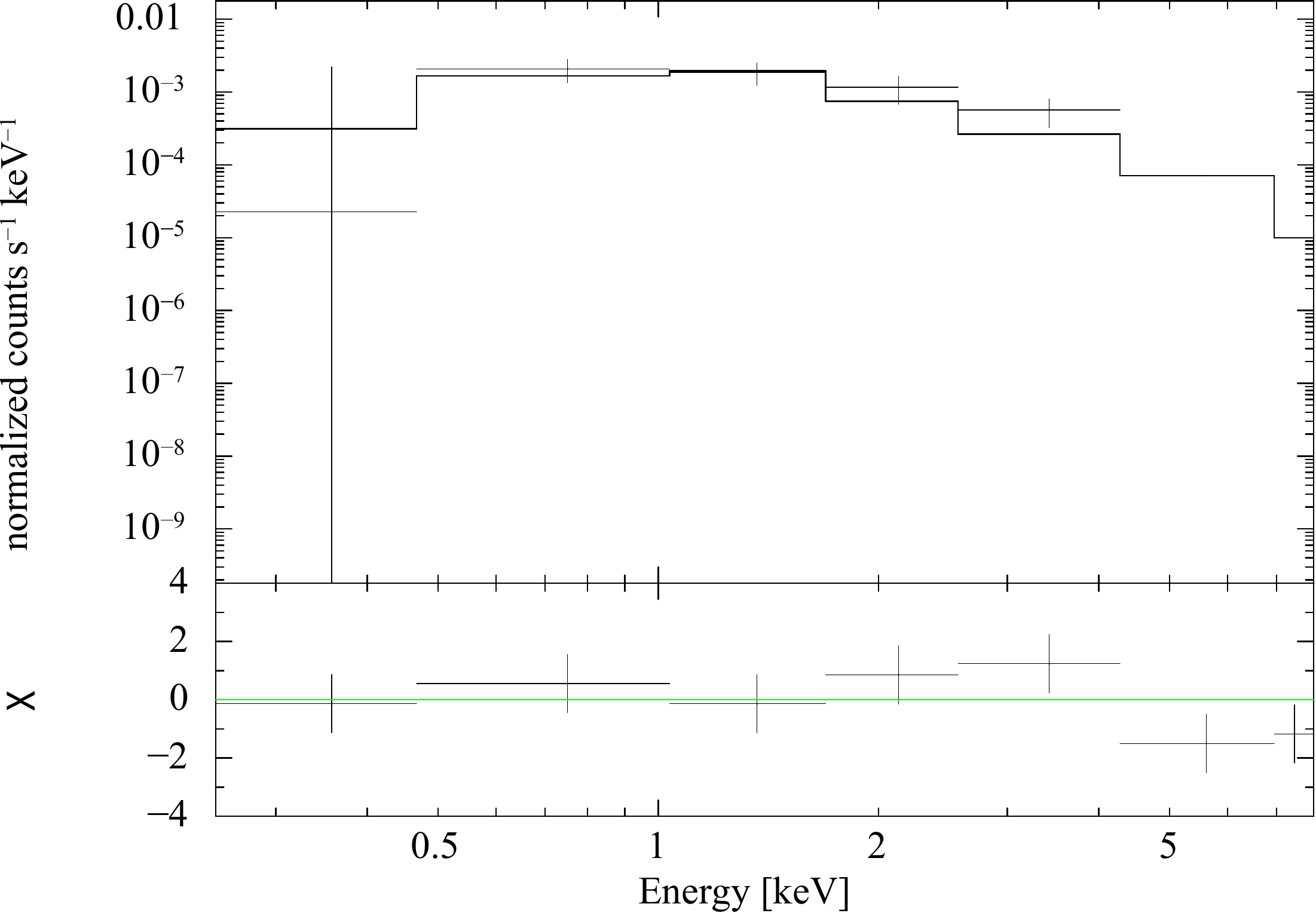}}
\end{minipage}
\caption{\chan\ spectrum of the trail-like nebula, 
the best-fit PL model and residuals.
}
\label{fig:tail_spec}
\end{figure}

For the spectral analysis of the trail-like nebula, we used only the \chan\ data set
since in the \xmm\ data it is somewhat blurred with several stars.
We extracted spectrum of the trail using the 30 arcsec $\times$ 300 arcsec rectangle region
shown in Fig.~\ref{fig:tail_reg} 
together with 
the regions used for the background.
It was binned to ensure at least 25 counts per energy bin and fitted in the 0.3--10 keV band.
The total number of counts in the spectrum is 811 while only 105 of them come from the source.
We tried the PL model and found that the column density is highly uncertain.
Thus, we fixed it at $3\times10^{20}$ cm$^{-2}$ which is compatible with all pulsar models (see Table~\ref{tab:psrpar}).
The resulting parameters $\Gamma = 1.9\pm0.5$, $K = 7.3^{+1.9}_{-1.8}\times 10^{-6}$ \phs,
the unabsorbed flux in the 
0.3--10 keV band $F_X=4.5^{+1.6}_{-1.1}\times10^{-14}$ \flux\ and $\chi^2_\nu=1.17$ (dof = 29).

Such spectrum can have a synchrotron nature.
The spectrum and the best-fit model are shown in Fig.~\ref{fig:tail_spec}.
The spectrum can also be equally well described ($\chi^2_\nu=1.16$, dof = 29) 
by the thermal bremsstrahlung model with a temperature $3.6^{+12.1}_{-1.9}$ keV and the unabsorbed flux in the 
0.3--10 keV band $F_X=3.7^{+1.8}_{-1.1}\times10^{-14}$ \flux.


\section{Discussion}
\label{sec:discussion}

\subsection{The pulsar spectral and timing properties}
\label{sec:discussion:spec}

The  time integrated X-ray spectrum of \psr\ the in 0.15--10 keV range can be well described by the composite model consisting of the thermal and PL components while the 
single PL model suggested previously by \citet{marelli2015} for a more narrow range of 0.3 -- 10 keV 
is statistically unacceptable in the extended range. 
In the case of the BB + PL model, the obtained effective temperature 
$T^{\infty}\approx54$~eV ($6.3\times 10^5$~K) is typical for the emission from the bulk of an NS surface but the  radius of the emitting area 
is smaller than an expected NS radius  of 10 -- 15 km (see Table~\ref{tab:psrpar}). 
The thermal emission could be produced by hot polar caps of the NS heated by relativistic particles from pulsar  magnetosphere.
For \psr, the `standard' pulsar polar cap radius is about 0.3 km \citep{sturrock1971}
which is compatible with the lower bound of the derived radius.
However, the derived temperature is too low for the polar cap emission 
\citep[cf.][]{potekhin2020}, rejecting this possibility. 
Thus, the BB + PL model might describe some hotter part of the pulsar 
surface while the other part is cooler and not observed in X-rays
as takes place, e.g., for the middle-aged PSR B1055$-$52 \citep*{mignani2010}.  
Note, however, that such interpretation implies that the obtained BB temperature
cannot be used for a comparison with the predictions of the cooling theory;
instead, the bolometric thermal luminosity
should be used for such a comparison 
(as discussed, e.g., in \citealt{potekhin2020}).
Remarkably, the best-fitting thermal luminosities given by the BB+PL and \textsc{nsmdip}+PL models
are compatible within uncertainties (see Table\ref{tab:psrpar}).

The \textsc{nsmdip}$+$PL models also give acceptable fits. Although they are possibly less preferable according to the
Bayesian criteria that disregard prior theoretical constraints on NS radii, 
they are more physically motivated.
In particular, they are based on the magnetic atmosphere models, computed
for realistic NS parameters, and they consistently
take into account the distributions of temperature and
magnetic field over the NS surface.
Combining the results from all these models (Table~\ref{tab:psrpar}), 
the estimated redshifted NS effective temperature $T^{\infty}\approx 20\pm4$ eV 
($0.23\pm 0.05$ MK). 
This makes the pulsar one of the coldest among all known NSs with estimated thermal luminosities, 
whose measured thermal emission from the surface is powered by cooling (see Section~\ref{sec:discussion:cool}).
We cannot constrain the pulsar viewing geometry (angles $\alpha$ and $\zeta$)  
from the time integrated 
spectra.

For the first time, we detected X-ray pulsations with the pulsar spin period.
The pulsations are significant only in the  soft band of 0.15--0.5 keV where the thermal emission component strongly dominates  in the spectrum of the pulsar.      
The pulse profile is a sine-like with a single pulse per period and the pulsed fraction of $\approx 18\pm6$ per cent, which is typical 
for  thermal emission from a bulk of the surface of NSs \citep[e.g.][]{pavlov&zavlin2000,zavlin2009}. This is independent  confirmation of the results of the spectral analysis. The pulsations can be due to nonuniform temperature distribution over the surface of the NS due to magnetic anisotropy  of the internal heat transfer to the star surface and the  magnetic beaming of the radiation in its atmosphere. Both factors are accounted in our \textsc{nsmdip} models, while only the first one can provide pulsations for the BB model which may reminiscent of the emission from a solid state surface of the NS. In any case, using the BB model with a single temperature is a very rude simplification at the non-uniform temperature distribution over the star surface.

\begin{figure*}
\includegraphics[width=\columnwidth]{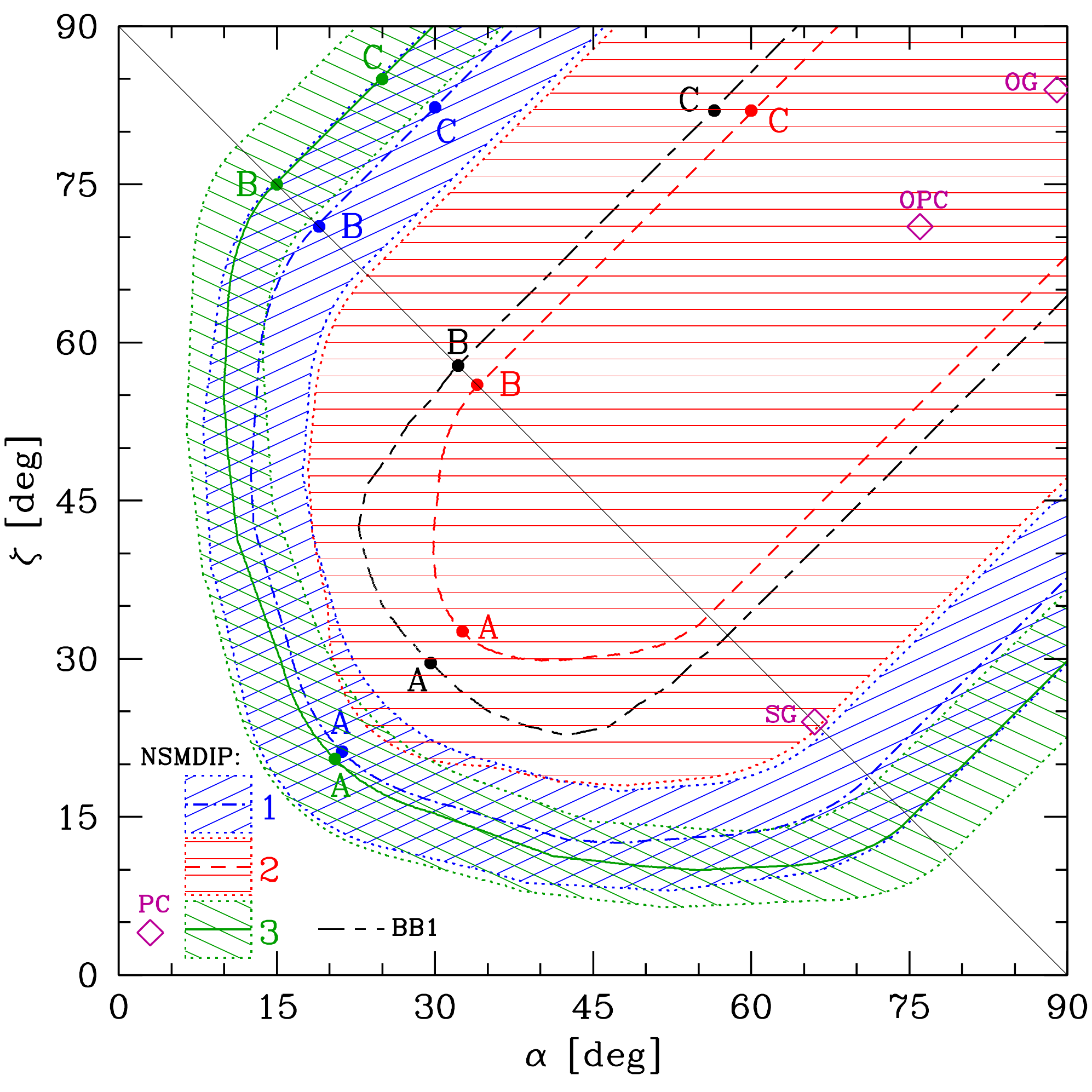} 
\includegraphics[width=\columnwidth]{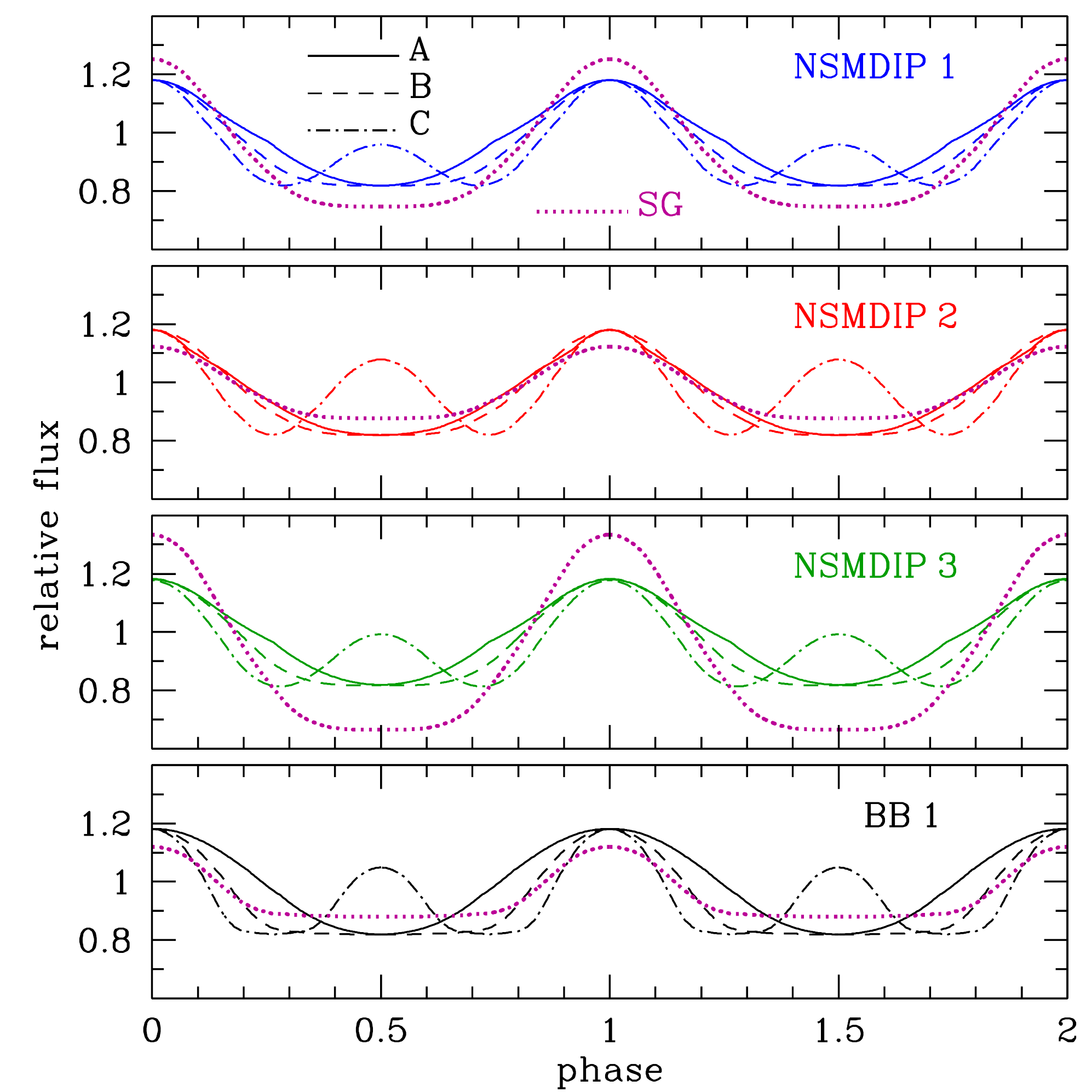} 
\caption{\textit{Left}: Regions of magnetic obliquity $\alpha$ and rotation axis inclination
$\zeta$ (hatched regions), compatible with the observed pulsed fraction $0.18\pm0.06$ of emission in the soft X-ray energy band 0.15--0.5 keV, 
and the $\alpha$ and $\zeta$ combinations (curves) that provide
the pulsed fraction 0.18, shown for the three \textsc{nsmdip}
atmosphere models according to the legend. For comparison,
a curve for a BB model (see text) is also plotted.
The diagonal line separates the parts of the hatched regions in the lower triangle, 
which are compatible with the singe-pulsed light curves, as
observed in Fig.~\ref{fig:prof}. The violet diamonds show 
the tentative $\alpha$ and $\zeta$ values obtained by
\citet{Pierbattista_15} using polar cap (PC), slot gap (SG),
outer gap (OG) and one pole caustic (OPC) models
of gamma-ray pulse formation.
\textit{Right}:
Light curves in the redshifted energy band 0.15--0.5 keV, computed
for models \textsc{nsmdip}\,1, 2 and 3 at $\lg T^\infty\mbox{(K)}=5.4$ 
(the three upper panels) and the
$\alpha$ and $\zeta$ angles shown by the respective points A, B and C in the left panel (solid, dashed,
and dot-dashed lines). 
For comparison, the bottom panel shows the light curves computed for the best-fit BB model. 
The violet dotted line in each panel shows the light curve
for $\alpha=66^\circ$ and $\zeta=24^\circ$
given by the fit  with 
the SG model 
\citep{Pierbattista_15}.
}
\label{fig:angles}
\end{figure*}

The observed soft X-ray pulsations can be used to constrain
the angles $\alpha$ and $\zeta$ that the rotation axis makes with the
magnetic axis and the line of sight (magnetic obliquity and pulsar inclination), 
assuming that the flux in the photon energy band
0.15--0.5 keV is dominated by thermal emission.
In the left panel of Fig.~\ref{fig:angles},  the hatched regions
correspond to the $\alpha$ and $\zeta$ values that provide the observed 
pulsed fractions of thermal radiation from 0.12 to 0.24
in the 0.15--0.5 keV energy band for the atmosphere models
\textsc{nsmdip}\,1--3, according to the legend. 
We see that for all the considered cases $\alpha+\zeta\ga30\degs$. 
Because of a higher redshift
in model \textsc{nsmdip}\,2, it shows a stronger gravitational light bending, 
which leads to a stronger smearing of the light curves, compared with the two other models.
Therefore, higher $\alpha$ and $\zeta$ are needed to obtain the same pulsed fraction.
In particular, the upper limit of 24 per cent is never reached 
(that is why the hatched area has a shape of a bell rather than a horseshoe in this case).
The absence of an inter-pulse on the observed light curve in Fig.~\ref{fig:prof} 
suggests that $\alpha+\zeta\lesssim90\degs$. 
The curves in  Fig.~\ref{fig:angles}, left,  show the combinations
of $\alpha$ and $\zeta$ that provide the pulsed fraction 18 per cent.
For comparison, we show an analogous line for the best-fitting BB
model from Table~\ref{tab:psrpar} ($R^\infty=2.2$ km), assuming 
that thermal radiation comes from two uniformly heated circular 
regions around the poles on the surface of a star with
the same $M=1.4\,\msun$ and $R=12.6$ km as in models $\textsc{nsmdip}$\,1 and 3.
In contrast to the atmosphere radiation, which is has a peaked 
angular distribution, the BB radiation obeys the Lambert's cosine law.
For this reason, the same pulsed fraction is reached for higher $\alpha+\zeta$ values.
For $\textsc{nsmdip}$\,2, the above-mentioned smearing of the light curves
due to the light bending effect is so strong that the pulsed fraction does not exceed 5 per cent, 
meaning that the BB model is inapplicable in this case.
We also show the tentative $\alpha$ and $\zeta$ values obtained by
\citet{Pierbattista_15}, who fitted geometrical models to the observed $\gamma$-ray
pulse profile, using different theoretical models of $\gamma$-ray pulse formation.
We see that one of the models (the slot gap model) is compatible 
with \textsc{nsmdip}\,1 and \textsc{nsmdip}\,2.
Examples of theoretical light curves, calculated as described in Appendix~\ref{appendix}, are shown in
the right panel of Fig.~\ref{fig:angles}.
For each model we show three cases, marked by letters A, B and C,
corresponding to the magnetic obliquity $\alpha$ and inclination $\zeta$ values
shown in the left panel. 
Note, that case C does not agree with the observed profile since it predicts the presence of an inter-pulse and thus can be excluded.

The low number of counts does not allow us to perform the phase-resolved spectral analysis 
which could help to distinguish between different geometries and spectral models. 
Deeper X-ray observations are necessary to do that. Ultraviolet (UV) observations 
could also clarify the situation whether the thermal emission is best descried by 
the \textsc{nsmdip} or by the BB (solid state) spectral models as they predict 
different fluxes in this range. 

Implementation of the extinction--distance relation in the fitting procedure
allowed us to estimate the distance to \psr.
We note, that due to the large relative uncertainties in this relation especially at low distances 
(see Appendix~\ref{appendix:nh-d}), the constraints on $D$ are rather weak especially 
for the BB + PL model where \nh\ uncertainties are larger than for the atmosphere models.
Thus, the BB + PL model gives the distance up to $\approx 1$ kpc while the {\sc nsmdip} + PL models 
resulted in the range of 0.1--0.4 kpc.
The former estimate is compatible with the `pseudo'-distance of 0.8 kpc \citep{marelli2015}.
Note,that the obtained distance range gives a reasonable $\gamma$-ray efficiency 
$\eta_\gamma = L_\gamma/\dot{E}$ of $\approx$0.006--0.6 where $L_\gamma = 4\pi D^2 G_{\rm 100}$ 
is the $\gamma$-ray luminosity and $G_{\rm 100}=2.6\times10^{-11}$ \flux\ \citep{marelli2015} is the $\gamma$-ray flux above 100 MeV.

As for the PL spectral component, the derived photon index range of 1.5--1.9 is typical 
for pulsars \citep[e.g.][]{kargaltsev&pavlov2008}.
For all the  models, the unabsorbed non-thermal flux in the 2--10 keV band is 
$F_X\approx2\times 10^{-14}$ \flux. 
This corresponds to X-ray luminosity $L_X$ of $\approx (0.24-3.83)\times 10^{29}$~erg s$^{-1}$ 
for the distance range of 100--400 pc provided by the atmosphere models and up to $2.4\times 10^{30}$~erg s$^{-1}$erg s$^{-1}$ with the upper bound on the distance of 1 kpc as provided by the BB model.
The corresponding X-ray efficiency of $\eta_X = L_X/\dot{E} = 10^{-5.34}-10^{-4.14}$ and up to $10^{-3.35}$, respectively.
These values are compatible with  empirical dependencies of pulsar X-ray nonthermal luminosity and efficiency vs. characteristic age   \citep*[see e.g.][]{zharikov2006,zharikov&mignani2013}. 
Upper limits on the pulsed fraction in harder bands of $\approx 40-60$ per cent do not give any additional 
informative constraints on the properties of the non-thermal X-ray emission from the pulsar magnetosphere.


\subsection{The nature of the trail-like nebula}
\label{sec:discussion:neb}

As was noted in Section~\ref{sec:data}, there is a thin straight feature protruding north-west from the pulsar (in equatorial coordinates; Fig.~\ref{fig:tail_reg}) 
likely associated with it. 
Its spectrum can be well described by the PL model with parameters typical for 
X-ray synchrotron nebulae powered by pulsars, known as PWNe.
Its length is $\approx 8$ arcmin which corresponds to $\approx0.23-2.3$ pc for the obtained distance range of 0.1--1 kpc (see Table~\ref{tab:psrpar}).
This is compatible with lengths of other PWNe \citep[e.g.][]{kargaltsev2017}.

We can assume that the \psr\ moves in the south-east direction 
and thus the feature is 
a trail-like PWN as observed e.g. for PSR J1741$-$2054 \citep{auchettl2015}. 
On the other hand, this can be a misaligned outflow which is not aligned with the pulsars's proper motion (p.m.) direction 
\citep[see e.g.][and references therein]{reynolds2017,Kargaltsev2017PWN-review}.
If so, we do not see any hint of a `normal' tail-like PWN
protruding behind the pulsar in X-ray images.
Torus or jet structures typical for PWNe of younger pulsars  are also not detected.
However, the situation can be similar to the Guitar nebula powered by PSR B2224+65:
the guitar-shaped bow-shock nebula was detected in H$\alpha$ while no head-tail PWN 
was found within the 
shock in X-rays \citep{reynolds2017}.
Instead, X-ray observations revealed a jet-like feature 
inclined by $\approx118$\degs\ to the p.m. direction of B2224+65.
For our pulsar, no H$\alpha$ emission was 
detected around it 
with the 3.6 m WIYN telescope at  
a 300~s exposure \citep{Halpha-psr-survey}. 
However, the exposure 
may be too short to detect a fainter bow-shock at a high galactic latitude 
of $\approx 11\degs$ where the density of the interstellar matter is small 
and deeper observations are needed.

The tail interpretation suggests that \psr\ should move towards 
the Galactic plane (Fig.~\ref{fig:images}) raising a question about 
its birth site somewhere in the Galactic halo. 
For the misaligned outflow, the direction of p.m. remains unclear.
In the latter case, 
we can assume that \psr\ was born in the Galactic disk since
this possibility is more plausible than the birth in the halo.
Due to the natal kick in  the supernova explosion 
it could achieve high velocities and move away from the disk.
In such a case, we can estimate the \psr\ p.m. perpendicular to the Galactic plane.
At the pulsar  galactic latitude of $\approx 11\degs$ 
the p.m. $\mu \approx 47\ t^{-1}_{\rm 840\ kyr}$ mas~yr$^{-1}$,
where $t_{\rm 840\ kyr}$ is the pulsar true age normalized 
to its characteristic age $t_\mathrm{c}$.
This corresponds to the transverse velocity $v \approx 220\ t^{-1}_{\rm 840\ kyr} D_{\rm kpc}$ km~s$^{-1}$ 
which is compatible with the pulsars velocity distribution \citep{hobbs2005}
for the estimated distance range even if the true age is $\sim$5 times smaller.

The nebula spectrum can be also
described by a thermal bremsstrahlung emission model 
with a temperature of $3.6^{+12.1}_{-1.9}$ keV. In this case the nebula
 emission comes from the
shocked ISM and
the pulsar 
p.m. must be aligned 
with the nebula axis. 
Such situation  appears
for the 
tail of PSR J0357+3205
\citep{marelli2013}.
However, due to low count statistics
it is hard to distinguish between the PL and the thermal models in our case.

Measurement of the pulsar's p.m. 
is necessary to understand the nature of the feature and deeper X-ray observations are needed to constrain its shape and spectral properties.


\subsection{\psr\ and the cooling theory}
\label{sec:discussion:cool}

\begin{figure}
\includegraphics[width=\columnwidth]{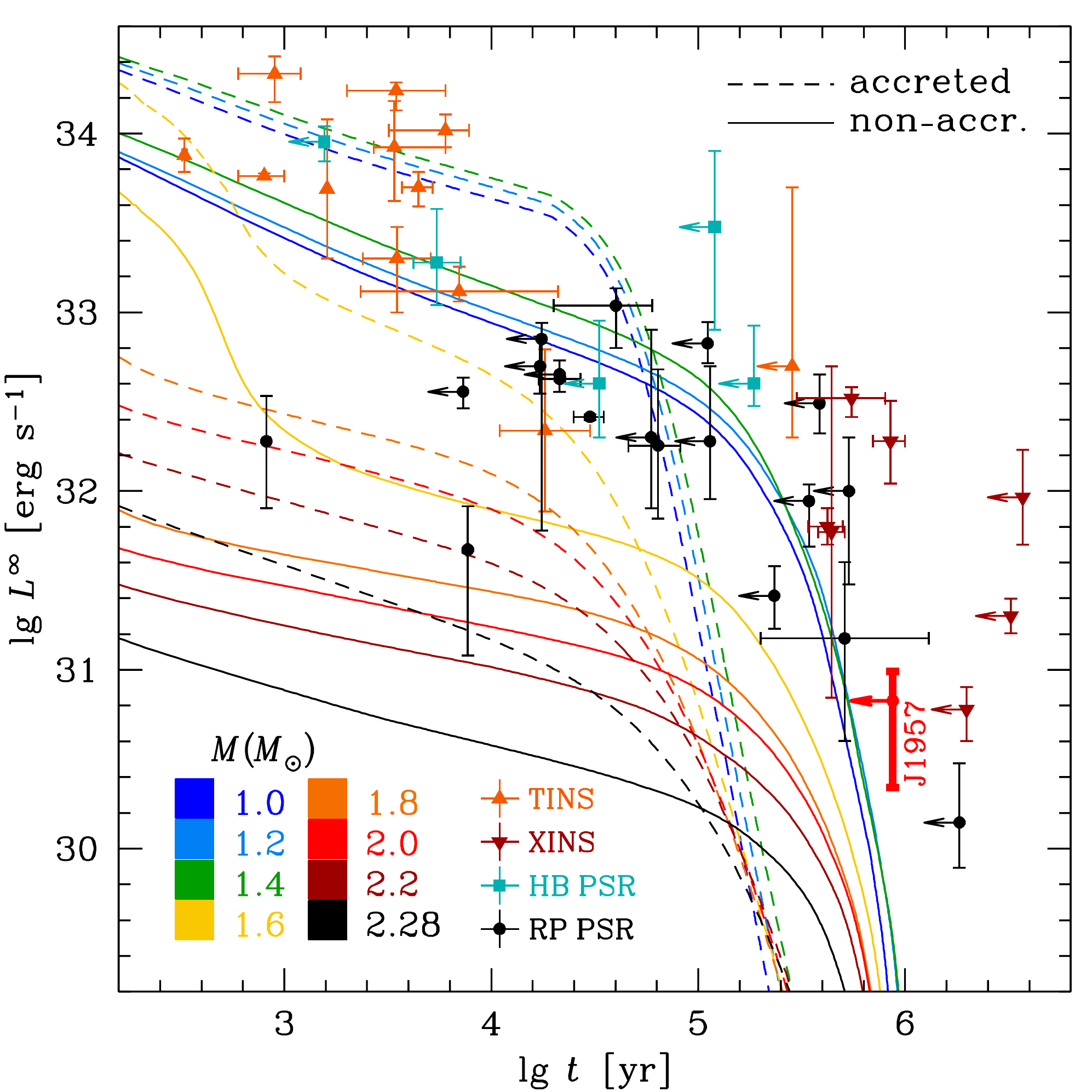} 
\caption{Cooling curves  of NSs with different masses (coded
with color), described by the EoS BSk24, for accreted and
non-accreted heat blanketing envelopes, compared with observations.
The data are plotted as indicated in the legend for NS classes
following \citet{potekhin2020}:  weakly magnetized thermally emitting
isolated NSs (TINS),  X-ray isolated NSs (XINS),
high-B pulsars (HB\,PSRs) and rotation powered (`ordinary') pulsars
(RP\,PSR). Vertical error bars show the
estimated uncertainties on bolometric thermal
luminosities, as seen by a distant observer. Horizontal error bars show
the estimated age intervals, whenever available; otherwise horizontal
arrows mark the characteristic ages (which are usually, although not
always, larger than the true ages). The position of \psr{} in this
figure corresponds to the best-fitting model {\sc nsmdip}\,1, while the error bar embraces
the models {\sc nsmdip}\,1--3 in Table~\ref{tab:psrpar}.
\label{fig:coolall}}
\end{figure}

\begin{figure*}
\includegraphics[width=.349\textwidth]{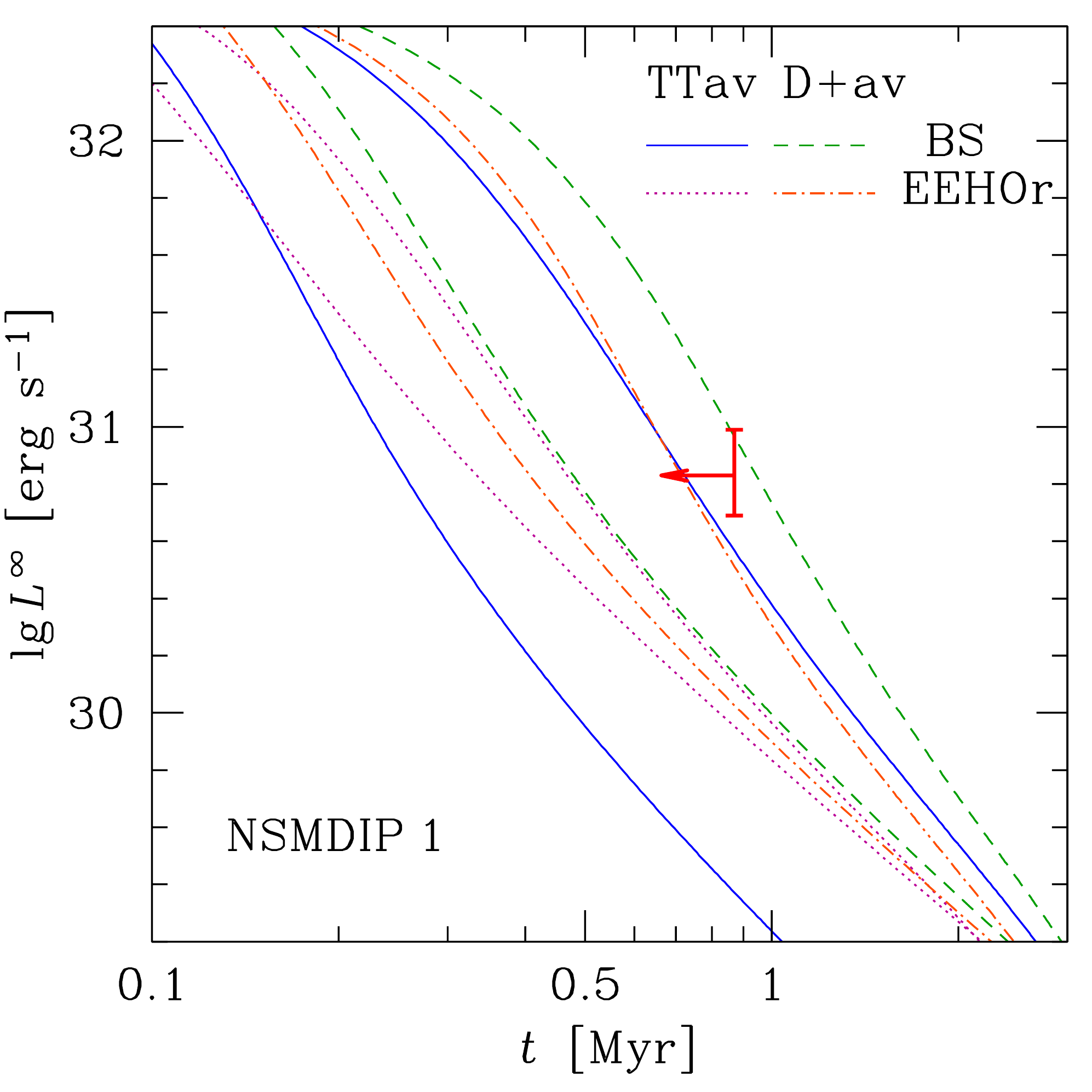} 
\includegraphics[width=.312\textwidth]{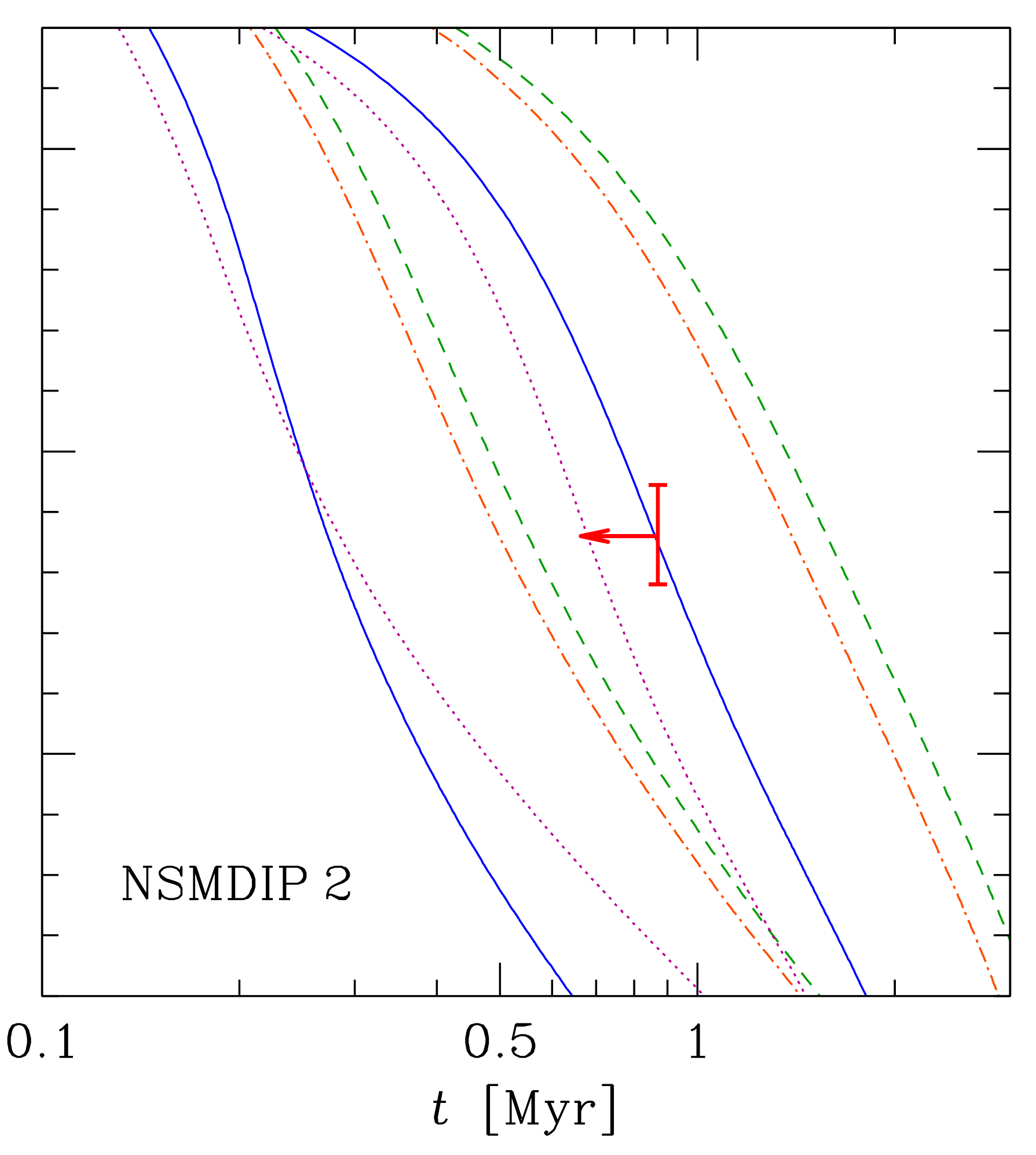} 
\includegraphics[width=.312\textwidth]{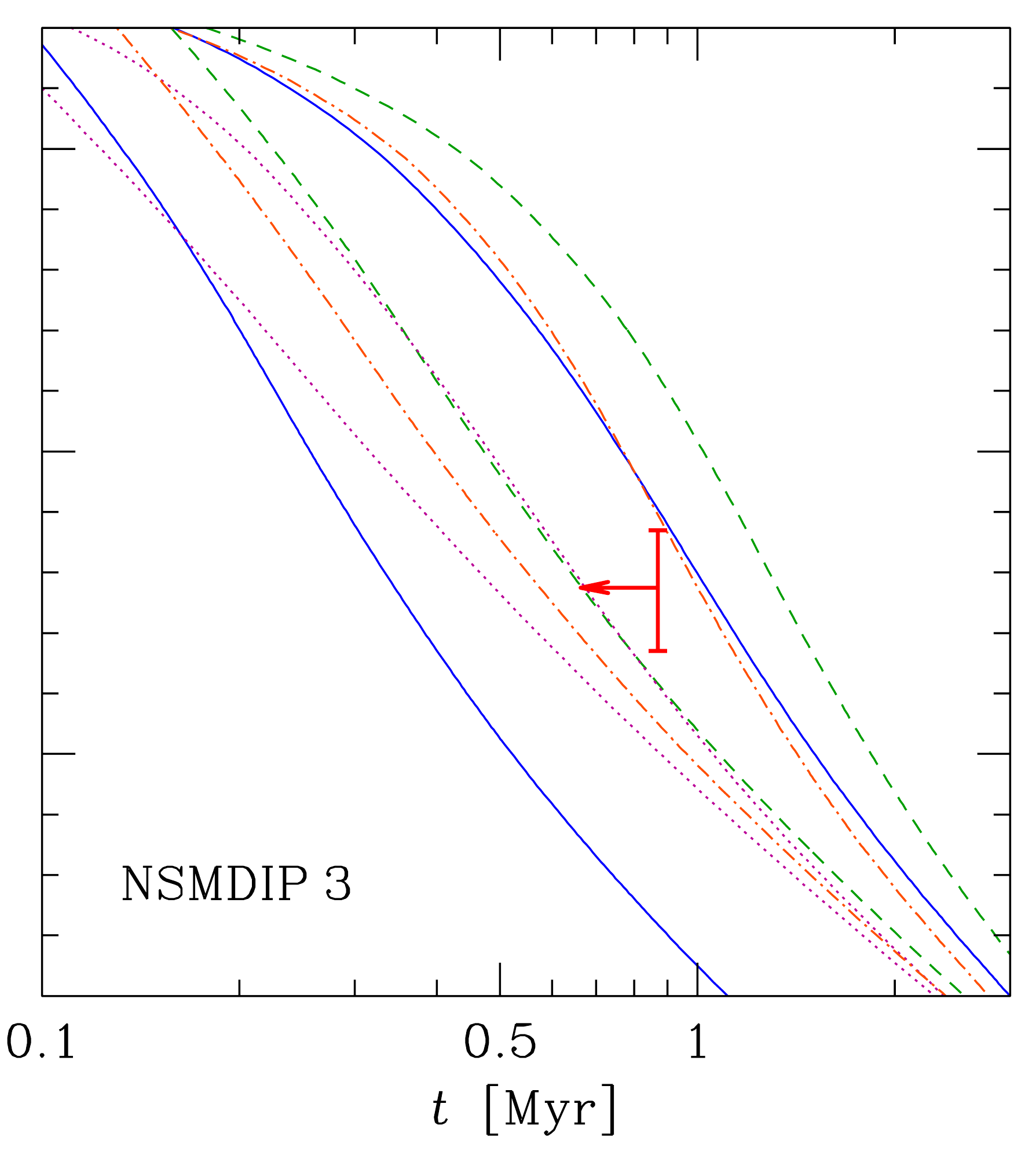} 
\caption{Late-time cooling of a strongly magnetized NS,
consistent with the spectral fit models \textsc{nsmdip}\,1, 2 and 3 (the left,
middle and right panels, respectively). 
The EoS model BSk24 with $M=1.4\,\msun$ is used for
\textsc{nsmdip} 1 and 3, and BSk26 with $M=2\,\msun$ 
is used for \textsc{nsmdip} 2. The polar magnetic field is
$B_\mathrm{p}=3\times10^{12}$~G (average $\langle B \rangle =
2.2\times10^{12}$~G) for \textsc{nsmdip} 1 and 2, while for \textsc{nsmdip} 3 we adopt
$B_\mathrm{p}=1.1\times10^{12}$~G ($\langle B \rangle =
7.9\times10^{11}$~G). Different line styles correspond to
different models of baryon superfluidity in the stellar core, as shown
in the legend: BS or EEHOr models for the proton pairing gap and
TTav or D+av models for the neutron triplet pairing (see text). In
each panel, the right/upper line of each type shows the cooling of a
NS covered by a non-accreted magnetized heat blanket made of
iron, while the left/lower line of each type corresponds to the fully
accreted heat-blanketing envelope composed of layers of hydrogen,
helium, carbon, and oxygen \citep{PYCG03,PC18}.
\label{fig:coolJ1957}}
\end{figure*}

As noted above, \psr\ can be one of the coldest cooling NSs with measured 
surface temperatures
(e.g., \citealt{potekhin2020} and references therein). 
In Fig.~\ref{fig:coolall} we compare the estimated thermal luminosities of different isolated NSs with
theoretical cooling curves (i.e., redshifted luminosities as functions of ages). 
For J1957 we use the luminosity estimates
reported in Table~\ref{tab:psrpar}. 
The error bar unites the uncertainties for the models \textsc{nsmdip}\,1--3. 
The model \textsc{nsmdip}\,1 is adopted as the best estimate, 
because it provides the lowest BIC value among these three models. 
The observational estimates for the other cooling NSs are taken from
\citet{potekhin2020}.\footnote{\url{http://www.ioffe.ru/astro/NSG/thermal/cooldat.html}}
For the NSs lacking timing-independent age estimates, including J1957, 
we plot $L^\infty$ against their characteristic ages $\tc$. 
In these cases the leftward arrows indicate that $\tc$ is likely to be larger than the true age, 
which is the common case, although there are exceptions where $\tc$ is somewhat lower than the true age 
(see, e.g., examples, discussion and references in \citealt{potekhin2020}). 
We see that \psr{} has the lowest thermal luminosity among all 
cooling NSs with ages $<1$ Myr. 

The theoretical cooling curves in Fig.~\ref{fig:coolall}
are calculated using the
numerical code presented by \citet{PC18}. The BSk24
model \citep{Pearson_18} is used for the composition and EoSp of the NS matter. The NSs are supposed to have either
non-accreted (ground state) heat blanketing envelopes or accreted
envelopes composed of helium, carbon, and oxygen up to the densities and temperatures
where these chemical elements can survive
\citep[see, e.g.,][]{PYCG03,PC18}.
The accreted envelopes are more heat-transparent than the ground-state ones. 
For this reason, the stars with the accreted envelopes are
brighter at the early stage of their evolution (at $t\lesssim10^4$ yr),
but they cool down faster and become colder at the late stage
($t\gtrsim10^5$ yr). An envelope may consist of the accreted material
only partially. In such cases the cooling rate is intermediate between
the non-accreted and fully accreted extremes shown in the figure.

The critical temperatures for singlet neutron superfluidity in the
inner crust and for proton and triplet neutron types of superfluidity
in the core of a NS are
evaluated, as functions of density, using the MSH, BS, and TTav
parametrizations of \citet{Ho_15}, which are based on theoretical models
computed, respectively, by \citet*{MargueronSH08},
\citet{BaldoSchulze07}, and \citet{TakatsukaTamagaki04}. 
As can be seen from Fig.~\ref{fig:coolall}, 
thermal luminosity of \psr\ is
higher than the predictions of the cooling model for the age $t=\tc$, so that
theoretical cooling curves pass to the left of the error bar in this figure.
However, as we mentioned above, it is likely that the true age $t$ is smaller than $\tc$.
If we treat $\tc$ as an upper limit to the true age,
than thermal luminosity of \psr\ is compatible with the considered theoretical model.
The smallest discrepancy between the best-fitting point and the theoretical
cooling curves in Fig.~\ref{fig:coolall} is observed for the model of an NS
with $M\approx 1.4\,\msun$, covered by the non-accreted heat-blanketing envelope.
In this case, an agreement between the model and observations
is reached, if we assume that $t\sim0.5\mbox{~Myr}\sim0.6\tc$.


However, the cooling NS models shown in Fig.~\ref{fig:coolall}, being
non-magnetic, are not fully
consistent with the models of strongly magnetized atmosphere spectra
(Appendix~\ref{appendix}). 
To produce cooling curves fully consistent with the spectral fitting, 
we employ a model of magnetized heat-blanketing
envelope with mass $10^{-5}\,\msun$. The bottom of such an envelope lies
in a deep layer of the outer crust (at densities $\sim10^{11}$ g
cm$^{-3}$), which is nearly isothermal at the considered 
NS ages. The interior of a NS
is treated as spherically symmetric, but 
temperature distribution in the envelope is essentially anisotropic. 
At the magnetic pole, the
mechanical and thermal structure of the envelope is computed
numerically, following \citet{PYCG03} and using updated microphysics
input as per \citet{PC18}. Thermal conductivity treatment
in the partially degenerate H and He
layers of the accreted envelope has been updated following \citet{Blouin_20}. 
The distribution of the effective temperature over the
surface is then taken consistent with the dipole
field, including the effects of General Relativity, as described in Appendix~\ref{appendix}. The interior of the star at densities $\gtrsim10^{11}$ g
cm$^{-3}$ is treated as spherically symmetric, with the microphysics (in
particular, thermal conductivities, synchrotron neutrino emission rates,
etc.) pertinent to the \emph{average} magnetic field strength for each
model.
In the cases \textsc{nsmdip} 1 and 3 ($M=1.4\,\msun$),
we use the composition and EoS model BSk24, while for
\textsc{nsmdip}\,2 we use BSk26 \citep{Pearson_18}. The latter EoS is softer 
and provides the core composition preventing neutron
star from the enhanced cooling at the assumed redshift $z_\mathrm{g}=0.44$,
which corresponds to $M\approx2\,\msun$. 

The results are shown in Fig.~\ref{fig:coolJ1957}. To test the effects of
baryon superfluidity we show, in addition to the models with the BS and TTav 
superfluidity of protons and neutrons (as in Fig.~\ref{fig:coolall}),
also the cooling curves computed using 
alternative proton and neutron superfluidity models in the core.
For an alternative proton superfluidity, we use the EEHOr 
parametrization by \citet{Ho_15} to 
the microscopic calculations of
proton critical temperature by
\citet{Elgaroy_96}, which predicted 
substantially stronger proton superfluidity than BS.
For an alternative superfluidity of neutrons,
we use the `Av18 SRC+P' model of \citet{Ding_16}.
which is marked `D+av' in Fig.~\ref{fig:coolJ1957}. 
In the latter case, the triplet neutron superfluidity is
suppressed due to the effects of many-body correlations (although
the maximum critical temperatures are similar in the TTav and D+av models,
the latter model predicts a narrower
range of densities where the neutrons are superfluid).

We see that the obtained estimates of thermal luminosity of \psr\ can
constrain the theoretical NS cooling models, constructed consistently
with the fitting. Assuming that the true
age of the pulsar does not exceed its characteristic age, we can
conclude that the NS model \textsc{nsmdip}\,1 is compatible
with all employed models of superfluidity and envelope composition,
but the NS models \textsc{nsmdip}\,2 and 3 are hardly compatible
with the suppressed neutron superfluidity (D+av), if the
envelope is non-accreted.
If we assume that the true age is close to $\tc$, then 
for each spectral model
we can select the best-fitting superfluidity and heat-blanketing envelope
models in Fig.~\ref{fig:coolJ1957}, for which the cooling curves 
are in a good agreement with the
spectral fitting results. This reinforces the suggestion that the
thermal-like part of the X-ray radiation of J1957 comes from the entire
surface and is powered by passive cooling.
Large uncertainties of the thermal luminosity provided by the BB model
do not allow us to constrain cooling models.


\section{Summary}
\label{sec:summary}

Using the \xmm\ and \chan\ observations of the middle-aged $\gamma$-ray pulsar \psr, 
we detected, for the first time, the thermal spectral component and X-ray pulsations.
We performed self-consistent modelling of thermal spectrum and cooling
for models of NSs with strong dipole magnetic field. 
We applied it to the observations and estimated the \psr\ thermal luminosity.
It is consistent with the NS cooling theory and provides certain constraints on the NS model parameters.
We found that \psr\ is one of the coldest middle-aged NSs with measured effective temperatures: 
its redshifted 
thermal luminosity is $L^\infty\approx(0.2-1.0)\times10^{30}$ erg s$^{-1}$.
This indicates  that \psr\  has already passed from a relatively slow neutrino cooling stage 
to a significantly faster photon stage. 
The  X-ray pulse-profile with a single pulse per period and with the pulse fraction of 18$\pm$6 per cent constraints 
  the pulsar viewing 
geometry  $30\degs\lesssim\alpha+\zeta\lesssim90\degs$. 

Using  the interstellar 
extinction-distance relation, 
we estimated the distance to the pulsar to be of $\approx$0.1-1 kpc.
Spectral fits with atmosphere models imply a most probable distance of 200 -- 300 ~pc.

We also detected a weak $\sim$8 arcmin long
trail-like feature connected with the pulsar. 
It most likely can be 
a PWN or a misaligned outflow 
powered by the pulsar.

Deeper X-ray observations are needed for more stringent constraints on the properties of the pulsar and the trail-like nebula.
Phase-resolved analysis of such data could be helpful to understand the pulsar geometry. 
Measurement of the pulsar's p.m. in X-rays  
is necessary to understand the nature of the nebula. 
Due to the low extinction and the likely proximity of \psr, it seems to be a good candidate for UV/optical observations.
This could confirm the low temperature and lead to more accurate distance estimate.

\section*{Acknowledgements}

We would like to thank the anonymous referee for useful comments and A.~A. Danilenko for helpful discussion.
The work was partially supported by 
the Russian Foundation for Basic Research (RFBR) according to the project 19-52-12013.
VFS thanks Deutsche  Forschungsgemeinschaft  (DFG) for financial support (grant WE 1312/53-1).  
His work was also partially  funded  by  the subsidy 0671-2020-0052 allocated to Kazan Federal University for 
the state assignment in  the sphere of scientific activities. 
DAZ thanks Pirinem School of Theoretical Physics for hospitality.
The scientific results reported in this article are based on observations obtained with \xmm, 
an ESA science mission with instruments and contributions directly funded by ESA Member States and NASA.

\section*{Data Availability}
The X-ray data are available through their respective data 
archives: https://www.cosmos.esa.int/web/xmm-newton/xsa for 
\xmm\ data and https://cxc.harvard.edu/cda/ for \chan\ data.



\bibliographystyle{mnras}
\bibliography{ms}


\appendix
\section{Atmosphere models for neutron stars
with dipole magnetic fields}
\label{appendix}

Magnetized plane-parallel NS atmosphere models are computed using an
advanced version of the code described in \citet*{SuleimanovPW09}. The
code has been modified to account for different inclinations $\theta_B$
of the magnetic field with respect to the local surface normal. 
Hydrogen composition is considered, taking into account
incomplete ionization at relatively low temperatures. 
The effects of the strong
magnetic field and the atomic thermal motion across the field on the
plasma opacities are treated
following \citet{PC03} with the improvements described in \citet*{PCH14}.
Polarization vectors and opacities of normal electromagnetic modes are
calculated as in \citet{Potekhin_04}. 

Physical models of emission from NSs should take into account magnetic
field and temperature distributions over the surface. We assume a
dipolar magnetic field, after accounting for the effect of General
Relativity, according to \citet{GinzburgOzernoy} (see also
\citealt{PavlovZavlin00}):
\begin{equation}
    B = B_\mathrm{p}\,\sqrt{\cos^2\gamma + f^2 \sin^2\gamma/4},
    \quad
    \cos\theta_B = (B_\mathrm{p}/B)\,\cos\gamma,
\label{Brel}
\end{equation}
where $B$ is the field strength, $\theta_B$ is the
field inclination to the surface normal at a magnetic colatitude 
$\gamma$,
\begin{equation}
    f =  \frac{2}{\sqrt{1-u}}\, \frac{u^2-2u-2(1-u)\,\ln(1-u)}{
      u^2+2u+2\,\ln(1-u)},
\label{frel}
\end{equation}
$u=r_\mathrm{g}/R$ is the compactness parameter,
$r_\mathrm{g}=2GM/c^2$ is
the gravitational radius, $G$ is the gravitational constant
and $c$ is the speed of light.
The distribution of local effective  temperature
$\Ts$ over the stellar surface is calculated using the results of
\citet{PYCG03}. In order to minimize model dependence, we assume the
$\Ts$-distribution of an iron heat-blanketing envelope. This
assumption does not change our results since, for any chemical
composition of the envelope, the dependence of $\Ts$ on $\theta_B$ is
similar to that given in \citet{GreensteinHartke}.

Model atmospheres for an inclined magnetic field require solving the
transfer problem in two dimensions. The optical properties of the magnetized
plasma depend on the angle $\eta$
between the photon wave vector $\bm{k}$
and the local magnetic
field $\bm{B}$. On the other hand, under the plane-parallel approximation
the radiation field naturally depends on the angles $(\theta_k,\phi_k)$,
where $\theta_k$ is the angle between $\bm{k}$
and the surface normal
and $\phi_k$ is the angle between the projections
of $\bm{k}$ and $\bm{B}$ on the surface.
To avoid interpolation of the opacities over such
a two-dimensional grid, the code solves the transfer problem over
an $(\eta,\psi)$ angular grid, where $\psi$ is
the azimuth associated to the polar angle $\eta$, and then
the transformation between the angular coordinates 
$(\eta,\psi)$ and $(\theta_k,\phi_k)$ is used
(see \citealt{Taverna_20} for more details):
\begin{eqnarray}
\cos\eta &=& \sin\theta_B\sin\theta_k\cos\phi_k +
        \cos\theta_B\cos\phi_k,
\\
\cos\psi &=& \frac{\cos\theta_k - \cos\eta\cos\theta_B}{
                |\sin\eta\sin\theta_B|}.
\end{eqnarray}

\begin{figure*}
\includegraphics[width=.36\textwidth]{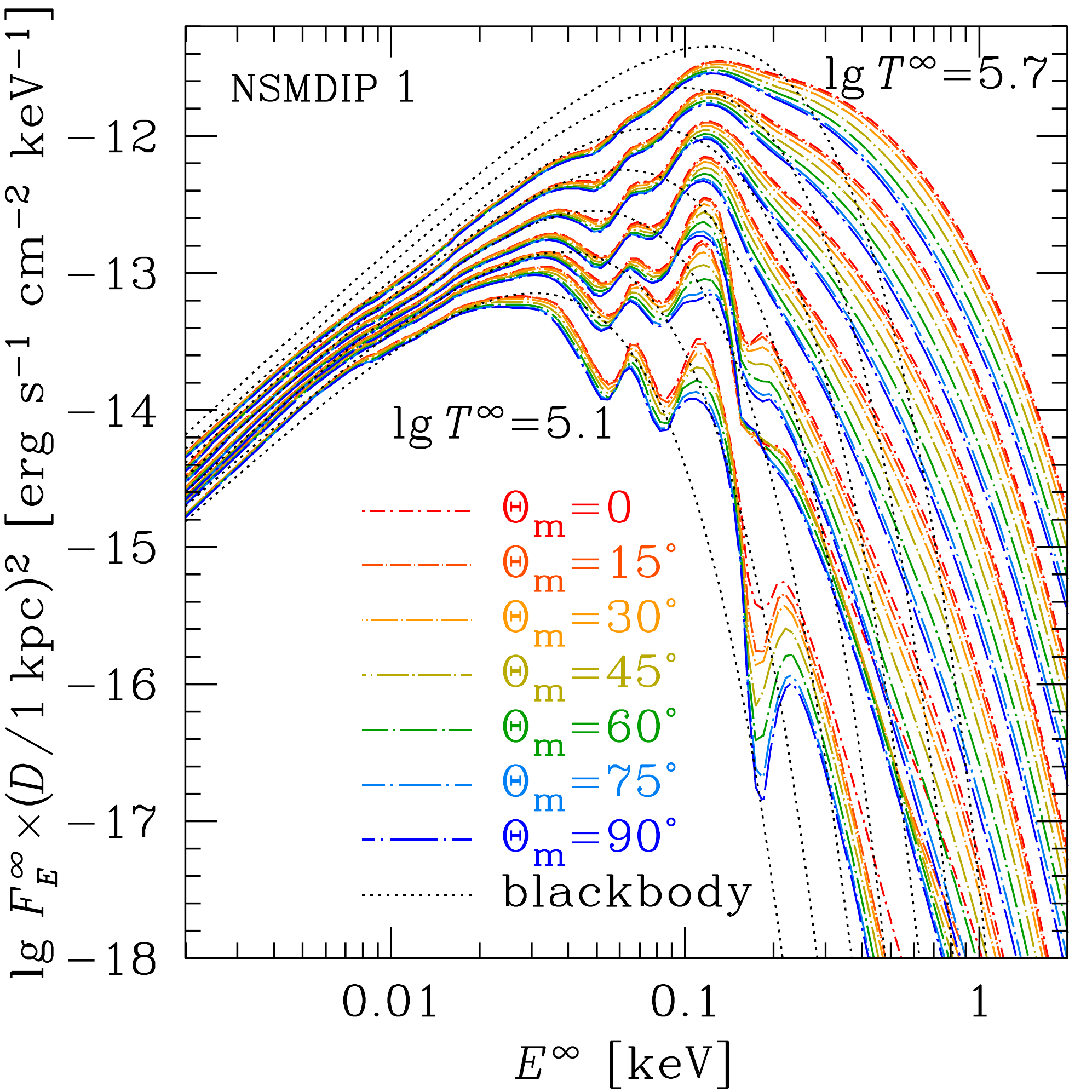} 
\includegraphics[width=.31\textwidth]{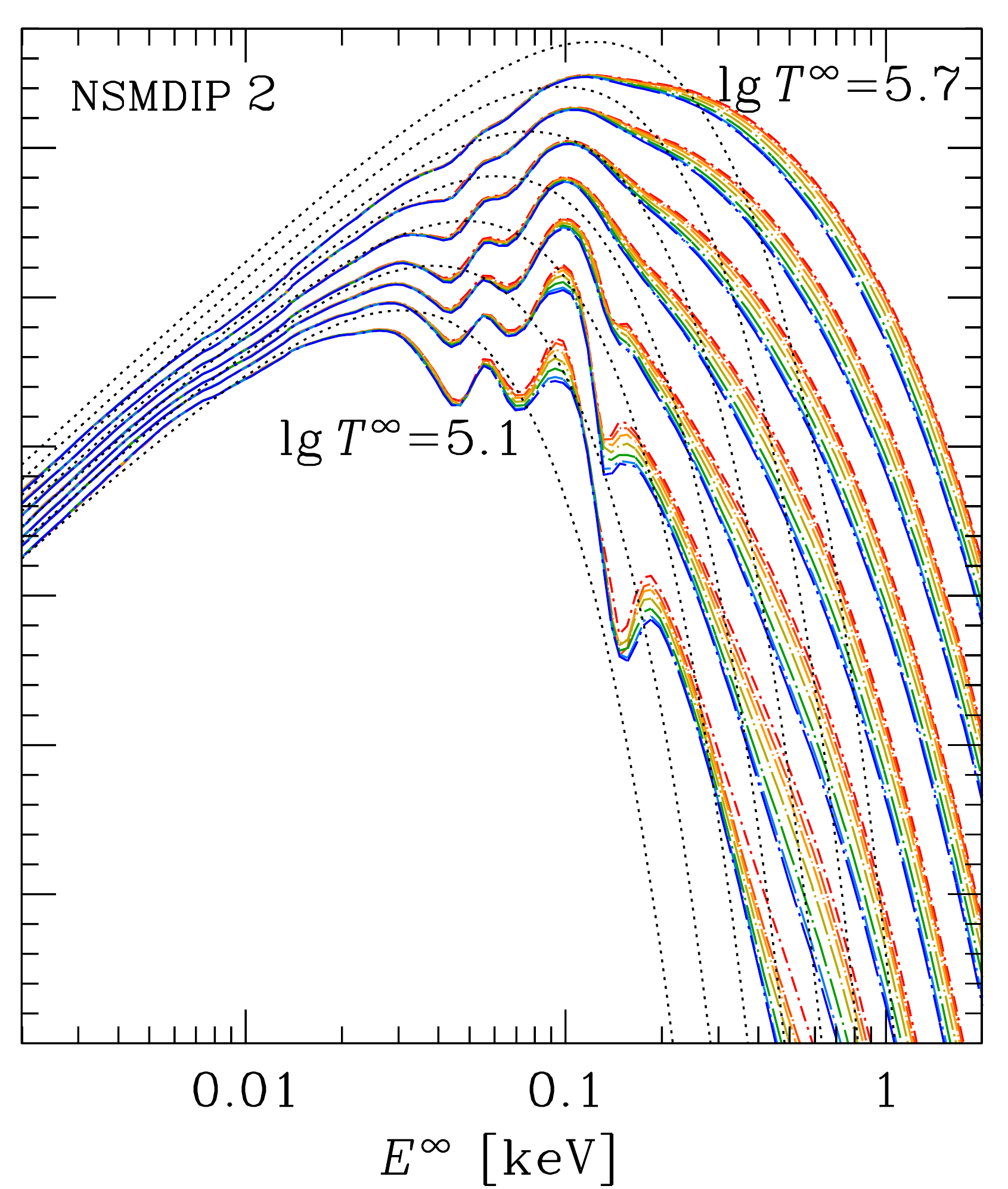} 
\includegraphics[width=.31\textwidth]{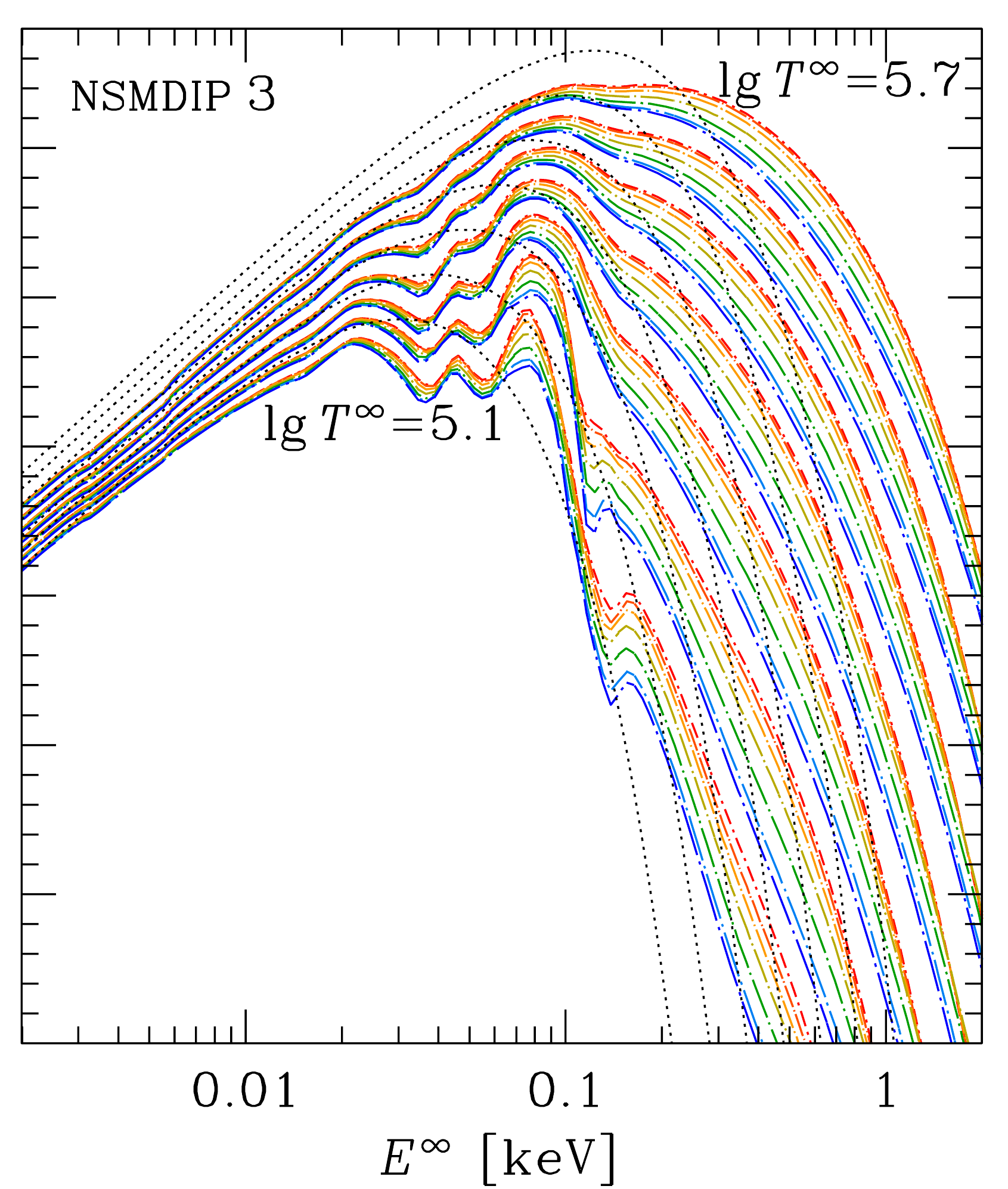} 
\caption{Unabsorbed redshifted spectral flux density as function of redshifted energy,
calculated according to
equation~(\ref{eq:flux1}) for models 1, 2 and 3 
listed in
Table~\ref{tab:dipset} (left, middle and 
right panels, respectively)
for different inclinations $\Theta_\mathrm{m}$ of magnetic dipole axis
(drawn with different line styles, according to the legend)
and redshifted effective temperatures ($\lg T^\infty$\,(K) = 5.1, 5.2, 5.3, 5.4, 5.5, 5.6 and 5.7, from bottom to top). For comparison,
the BB spectra at the same temperatures are shown by dotted lines.
\label{fig:spdip}}
\end{figure*}

The photon wave-vector at infinity $\bm{k}'$ differs from $\bm{k}$ at
the surface due to gravitational redshift and light-bending
\citep*{PechenickFC83,PavlovZavlin00}. The photon energy at infinity is
smaller than the photon energy at the surface by factor $\sqrt{1 -
u}=1/(1+z_\mathrm{g})$, where $z_\mathrm{g}$ is gravitational redshift. The
period of PSR J1957+5033 is sufficiently long so that we can assume that
the stellar surface is  spherical and neglect the effects of rotation
(see, e.g., \citealt{PoutanenBeloborodov06} for description of these
effects). Let us consider a surface element
d$S=R^2\,\mathrm{d}\cos\vartheta\,\mathrm{d}\varphi$, where $\vartheta$
and $\varphi$ are the polar and azimuthal angles. 
For the relation between $\vartheta$ and $\theta_k$
we use approximation \citep{Beloborodov02}
\begin{equation}
1 - \cos\vartheta = (1-\cos\theta_k)/(1-u).
\end{equation}
The flux observed from
this surface element is proportional to the specific intensity
$I_E^\infty$ and the solid angle $\mathrm{d}\Omega$ occupied by this
element on the observer's sky. This solid angle equals
\citep{Beloborodov02}
\begin{equation}
\mathrm{d}\Omega = \frac{\mathrm{d}S \cos\theta_k}{D^2}
\,\frac{1}{1-u}
\,\frac{\mathrm{d}\cos\theta_k}{\mathrm{d}\cos\theta_k'},
\label{dOmega}
\end{equation}
where $\theta_k'$ is the angle between the
local normal to the surface and the photon momentum 
at infinity.
Without loss of generality we can choose the polar coordinate axis along
the line of sight; then $\theta_k'=\vartheta$.
Since $I_E/E^3$ is invariant (\citealt*{MisnerThorneWheeler},
Sect.~22.6), the observed specific intensity $I_E^\infty$
is related to the emitted  intensity $I_E$ by a constant redshift factor
$I_E^\infty=I_E/(1+z_\mathrm{g})^3$.
Thus the flux is
\begin{equation}
\mathrm{d}F_E^\infty = I_E^\infty \mathrm{d}\Omega
 = \frac{R^2\cos\theta_k}{D^2}
 \frac{I_E}{(1+z_\mathrm{g})^3}
 \,\mathrm{d}\varphi\,\mathrm{d}\cos\vartheta.
\end{equation}
The monochromatic spectral flux density is then computed 
by integrating
the emission from different local patches over the stellar
surface. Making use of equation~(\ref{dOmega}), we obtain\footnote{This
is equivalent to equation~(8) of \citet{ho2008},
where we have restored the missed factor $\cos\theta_k$.}
\begin{equation}
F_E^\infty = \frac{R^2}{D^2(1+z_\mathrm{g})}
 \int_0^{2\pi}\!\! \mathrm{d}\varphi
 \int_0^{\pi/2}\!\!  I_E(\theta_k,\phi_k)
 \cos\theta_k \sin\theta_k \mathrm{d}\theta_k .
\label{eq:flux1}
\end{equation}
For an axisymmetric magnetic field,
the angle $\phi_k$ and 
the magnetic colatitude $\gamma$ are determined at every $\theta_k$ and
$\varphi$ by the relations
(cf.\ \citealt{ho2008})
\begin{eqnarray}
\cos\gamma &=& \cos\varphi\sin\vartheta\sin\Theta_\mathrm{m}
 + \cos\vartheta\cos\Theta_\mathrm{m}, 
\\
\cos\phi_k &=& \frac{\cos\gamma\cos\vartheta - \cos\Theta_\mathrm{m}
         }{
         |\sin\gamma\sin\vartheta|
         },
\label{eq:sinphik}
\end{eqnarray}
where $\Theta_\mathrm{m}$ is the angle between the magnetic axis and line of
sight.
The integration in equation~(\ref{eq:flux1})
is restricted by those angles $\theta_k$ that correspond to 
real values of $\vartheta$.

\begin{table}
	\centering
\caption{Parameter sets used in calculation of atmosphere models:
cosines of magnetic colatitude $\gamma$ and magnetic field inclination
$\theta_B$, logarithms of magnetic field strength $B$ and
local effective temperatures $\Ts$.}
\label{tab:dipset}
\begin{tabular}{l @{\hspace{1ex}} l @{\hspace{1ex}} l @{\hspace{1ex}} l @{\hspace{1ex}} l @{\hspace{1ex}} l @{\hspace{1ex}} l @{\hspace{1ex}} l @{\hspace{1ex}} l @{\hspace{1ex}} l}
\hline
\multicolumn{3}{l}{$\lg T^\infty$ (K)} & 5.1 & 5.2 & 5.3 & 5.4 & 5.5 & 5.6 & 5.7 \\
\hline
$\cos\gamma$ & $\cos\theta_B$ & $\lg B$ (G) & \multicolumn{7}{c}{$\lg\Ts$ (K)} \\
\hline
\multicolumn{10}{l}{\textsc{nsmdip} 1: $M=1.4\,\msun$, $R=12.6$~km ($z_\mathrm{g}=0.22$), $B_\mathrm{p}=3\times10^{12}$~G} \\
0.25 & 0.424 & 12.25 & 5.075 & 5.183 & 5.285 & 5.383 & 5.483 & 5.583 & 5.684 \\
0.50 & 0.724 & 12.32 & 5.188 & 5.292 & 5.393 & 5.492 & 5.591 & 5.691 & 5.791 \\
0.75 & 0.900 & 12.40 & 5.238 & 5.337 & 5.437 & 5.537 & 5.637 & 5.737 & 5.837 \\
1    & 1     & 12.48 & 5.266 & 5.361 & 5.459 & 5.560 & 5.661 & 5.762 & 5.861 \\
\multicolumn{10}{l}{\textsc{nsmdip} 2: $M=2\,\msun$, $R=11.4$~km ($z_\mathrm{g}=0.44$), $B_\mathrm{p}=3\times10^{12}$~G} \\
0.25 & 0.398 & 12.28 & 5.139 & 5.246 & 5.348 & 5.447 & 5.546 & 5.647 & 5.748 \\
0.44 & 0.633 & 12.32 & 5.236 & 5.340 & 5.442 & 5.541 & 5.640 & 5.740 & 5.840 \\
0.73 & 0.875 & 12.40 & 5.308 & 5.408 & 5.508 & 5.608 & 5.708 & 5.807 & 5.907 \\
1    & 1     & 12.48 & 5.341 & 5.437 & 5.535 & 5.636 & 5.737 & 5.837 & 5.937 \\
\multicolumn{10}{l}{\textsc{nsmdip} 3: $M=1.4\,\msun$, $R=12.6$~km ($z_\mathrm{g}=0.22$), $B_\mathrm{p}=1.1\times10^{12}$~G} \\
0.19 & 0.335 & 11.80 & 5.037 & 5.137 & 5.236 & 5.337 & 5.439 & 5.542 & 5.645 \\
0.56 & 0.775 & 11.90 & 5.206 & 5.306 & 5.405 & 5.505 & 5.605 & 5.705 & 5.805 \\
0.87 & 0.953 & 12.00 & 5.249 & 5.349 & 5.450 & 5.550 & 5.650 & 5.749 & 5.848 \\
1    & 1     & 12.04 & 5.260 & 5.359 & 5.461 & 5.561 & 5.661 & 5.760 & 5.859 \\
\hline
\end{tabular}
\end{table}

For each neutron star parameter set, the local plane-parallel atmosphere
models are computed at the magnetic pole and at three
magnetic latitudes between the pole and the equator, according to 
Table~\ref{tab:dipset}. The atmosphere
model at the equator needs not to be calculated, because temperature
at low latitudes is so low that it cannot noticeably affect
the observed flux (in
practice, we use the blackbody spectrum at the equator, 
but we have checked
that with alternative models $F_E^\infty$ remains the same within 3 per cents). 
At each fixed magnetic colatitude $\gamma$,
the opacities, polarizabilities,
and EoS of the hydrogen plasma were computed on a fixed
grid of plasma temperature and density, from which the values required
during the radiative-transfer calculation were obtained by interpolation.

To calculate the integral (\ref{eq:flux1}), the specific intensity $I_E$
is evaluated at arbitrary $\gamma$, $E$,
$\theta_k$ and $\phi_k$ from the computed values by interpolation. It
should be noted that $B$-dependent absorption features  would produce
series of lines, if we kept $E$ fixed during this
interpolation. In reality, such features are broadened due
to the smooth variation of $B$ with $\gamma$. In order to reproduce this
broadening and thus get rid of the non-physical series of lines, we first remap our calculated $I_E$ as
a function of ratio $E/B$ and then interpolate it in $\theta_k$,
$\phi_k$, and $B(\gamma)$ for every fixed $E/B$ (cf.\ \citealt{ho2008}).
The result of such integration is shown in Fig.~\ref{fig:spdip}.

The local effective temperature $T_\mathrm{s}$ and the 
global effective temperature $\Teff$ are defined by the Stefan-Boltzmann law
\begin{equation}
    \sigma_\mathrm{SB}\Ts^4 = F_r,
    \quad
    4\pi R^2 \sigma_\mathrm{SB}\Teff^4 = L_r
\end{equation}
where
\begin{equation}
   F_r = \int_0^\infty \mathrm{d}E
 \int_0^{2\pi}\mathrm{d}\phi_k \int_0^{\pi/2}
 I_E(\theta_k,\phi_k)
 \cos\theta_k\sin\theta_k\, \mathrm{d}\theta_k
\end{equation}
is the local flux density, which depends on
the magnetic colatitude $\gamma$, and
\begin{equation}
    L_r= R^2 \int_0^{2\pi}\mathrm{d}\varphi
    \int_0^{\pi}\,
    F_r(\gamma)\,\sin\vartheta\,\mathrm{d}\vartheta
\end{equation}
is the local bolometric luminosity.
The redshifted (`apparent') luminosity, effective temperature and
radius as detected by a distant observer are
(e.g., \citealt{Thorne77})
\begin{eqnarray}
&&   L^\infty = L_r /(1+z_\mathrm{g})^2 =
     4 \pi \sigma_\mathrm{SB} (T^\infty)^{4} (R^\infty)^2,
\label{srt-L_gamma_infty}\\
&&   T^\infty = \Teff /(1+z_\mathrm{g}), \quad
     R^\infty = R(1+z_\mathrm{g}).
\label{therm-T_s_infty}
\end{eqnarray}

For each model (\textsc{nsmdip} 1, 2 or 3),
we fix radius $R$, redshift $z_\mathrm{g}$ {and $B_\mathrm{p}$} to the values indicated 
in Table~\ref{tab:dipset} and treat the effective temperature $T^\infty$,
magnetic axis inclination $\Theta_\mathrm{m}$ and distance $D$ as
continuous  adjustable parameters {to fit the observed spectral fluxes using calculated grids  of $F_E^\infty$ }.  

In the axisymmetric model, the pulsar geometry is determined by the angles $\alpha$ and $\zeta$ that the spin axis makes
with the magnetic axis and with the line of sight, respectively \citep[e.g.,][]{PavlovZavlin00}.
To produce phase-resolved spectra, it is sufficient to calculate 
\begin{equation}
    \cos\Theta_\mathrm{m} = \sin\zeta\sin\alpha\cos\phi
     + \cos\alpha\cos\zeta
\end{equation} 
for each rotation phase $\phi$. 
Some light curves computed by integration of such phase-resolved spectra
are shown in Fig.~\ref{fig:angles}.


\section{Interstellar absorption--distance relation}
\label{appendix:nh-d}

\begin{figure}
\begin{minipage}[h]{1.\linewidth}
\center{\includegraphics[width=1.\linewidth,clip]{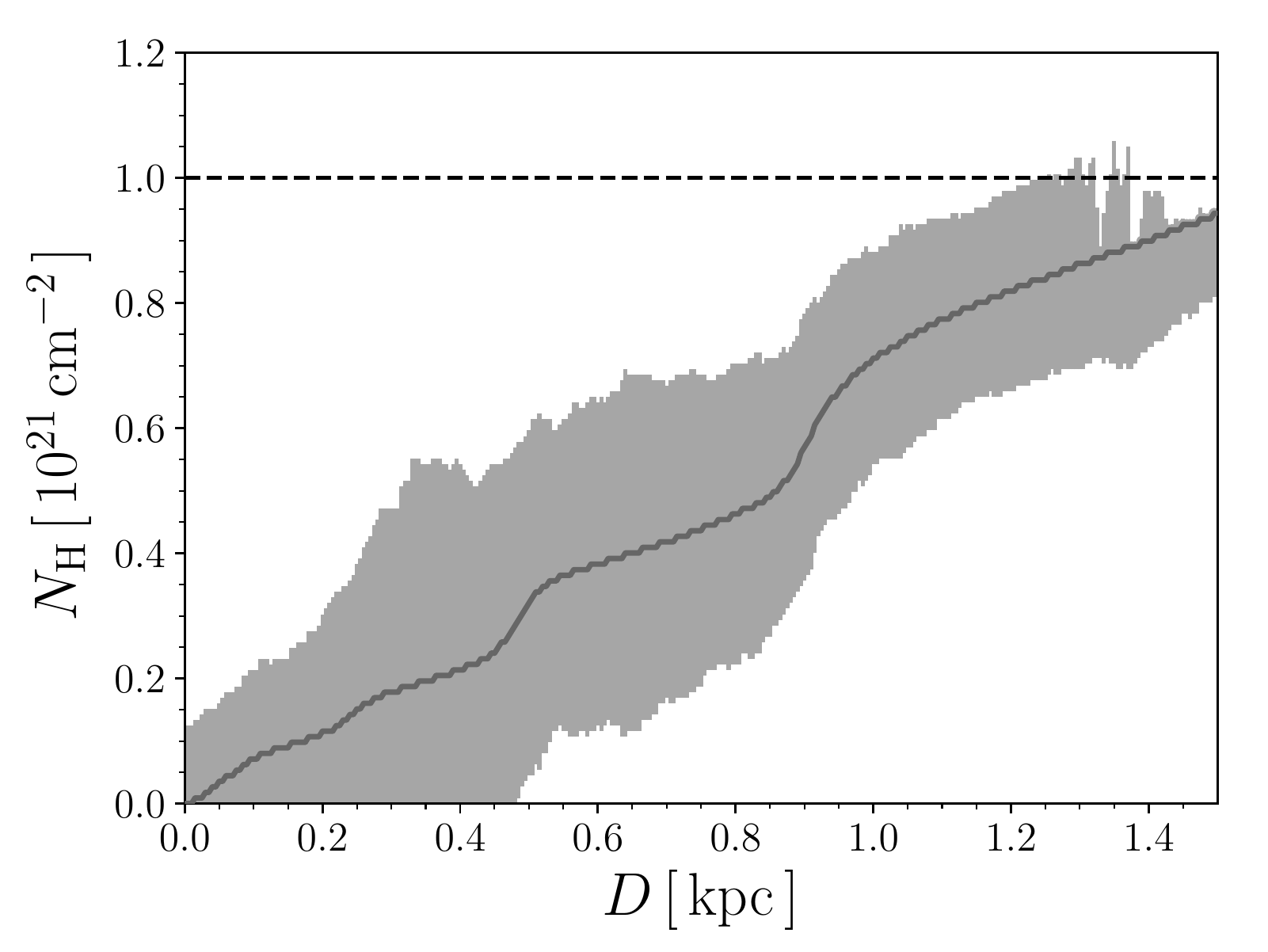}}
\end{minipage}
\caption{The relation between the interstellar absorption \nh\ and the distance $D$ (the dark gray line) with uncertainties shown in light gray 
in the direction towards \psr.
The dashed black line shows the maximum \nh\ derived from the spectral analysis of two AGNi in the pulsar field (see text for details).}
\label{fig:nh-d}
\end{figure}

We estimated the distance to \psr\ including the interstellar absorption--distance relation shown 
in Fig.~\ref{fig:nh-d} as a prior in the fitting procedure.
The relation was derived in the following way.
We used the 3D map of the local interstellar medium presented in \url{https://stilism.obspm.fr/} 
(see \citealt{lallement2014, capitanio2017, lallement2018} for details) 
to obtain the relation between the interstellar extinction $E(B-V)$ and the distance $D$ in the pulsar direction. 
Then $E(B-V)$ was converted to the X-ray absorbing column density \nh\ utilising the relation by \citet{foight2016}.
We used linear interpolation to derive the \nh--$D$ dependence between the obtained points.

We also independently  estimated the maximum absorption in the pulsar direction using two 
brightest extra-galactic  X-ray sources in the \psr\ field 
with coordinates R.A., Dec. = (19\h57\m16\fss491, +50\degs40\amin17\farcs280) and 
R.A., Dec. = (19\h56\m48\fss205, +50\degs39\amin32\farcs101). 
They  have optical counterparts 
in the Pan-STARRS \citep{ps2016} and WISE \citep{wise2014,unWISE2019} catalogues and are detected with the optical 
monitor (OM) on-board \xmm. 
According to their spectral energy distributions, the sources are  active galactic nuclei (AGNi).
We extracted their X-ray spectra, grouped them to ensure at least 25 counts per energy bin
and fitted with the absorbed model for AGN \textsc{optxagn}. 
The resulting column density \nh\ shown by the dashed black line in Fig.~\ref{fig:nh-d} is about 10$^{21}$ cm$^{-2}$
which is in agreement with the value obtained from the $E(B-V)$--$D$ relation.

The spectral models 
which we used to describe the pulsar thermal emission
includes the ratio of the emitting area radius and the distance as a parameter.
Implementation of the \nh--$D$ relation
allowed us to separate it into two independent parameters.





\bsp	
\label{lastpage}
\end{document}